\documentclass[floats,floatfix,amssymb,prd,twocolumn,superscriptaddress,nofootinbib,preprintnumbers]{revtex4-1}

\usepackage{subcaption,ragged2e}
\DeclareCaptionJustification{justified}{\justifying}
\usepackage{amssymb,amsmath,verbatim,mathtools,needspace,enumitem,etoolbox,graphicx,microtype,afterpage,bm}
\usepackage{tikz}
\usetikzlibrary{shapes.misc, positioning, arrows.meta}
\usepackage{framed}
\usepackage{soul}
\usepackage{hyperref}
\hypersetup{colorlinks, citecolor=bluscuro, linkcolor=black, urlcolor=bluscuro}
\definecolor{rossos}{cmyk}{0,1,1,0.55}
\definecolor{bluscuro}{rgb}{0.15, 0.2, .85}
\definecolor{bluchiaro}{cmyk}{1,.3,0.,0.1}
\definecolor{ForestGreen}{rgb}{0.13, 0.55, 0.13}
\definecolor{TLGreen}{RGB}{50, 164, 49}
\definecolor{TLOrange}{RGB}{231,180,22}
\definecolor{TLRed}{RGB}{204,50,50}

\usepackage{multirow,pifont,lmodern,float,tabularx,booktabs}
\usepackage[all]{hypcap}
\usepackage[T1]{fontenc}
\usepackage[utf8]{inputenc}
\usepackage{cleveref}

\allowdisplaybreaks

\newcommand{\Np}{N_{\rm p}}
\newcommand{\OGW}{\Omega_{\rm GW}}
\newcommand{\aPL}{\alpha_{\rm PL}}
\newcommand{\Az}{A_\zeta}
\newcommand{\be}{\begin{equation}}
\newcommand{\ee}{\end{equation}}

\begin{document}

\title{
Forecasting Constraints on SIGW with Future Pulsar Timing Array Observations
}

\author{Chiara Cecchini}
\email{chiara.cecchini@unitn.it}
\affiliation{Department of Physics, University of Trento, Via Sommarive 14, 38122 Povo (TN), Italy}
\affiliation{Trento Institute for Fundamental Physics and Applications-INFN, Via Sommarive 14, 38122 Povo (TN), Italy}
\affiliation{CERN, Theoretical Physics Department,
Esplanade des Particules 1, Geneva 1211, Switzerland}

\author{Gabriele Franciolini}
\email{gabriele.franciolini@cern.ch}
\affiliation{CERN, Theoretical Physics Department,
Esplanade des Particules 1, Geneva 1211, Switzerland} 

\author{Mauro Pieroni}
\email{mauro.pieroni@cern.ch}
\affiliation{CERN, Theoretical Physics Department,
Esplanade des Particules 1, Geneva 1211, Switzerland}

\date{\today}

\begin{abstract}
Pulsar Timing Arrays are playing a crucial role in the ongoing gravitational wave astronomy revolution. 
The current evidence for a stochastic gravitational wave background (SGWB) at nHz frequencies offers an opportunity to discover cosmological signals and threatens the observability of other subdominant GWs. 
We explore prospects to constrain second-order scalar-induced GWs (SIGWs) associated with enhanced curvature perturbations in the primordial universe, forecasting realistic future PTA datasets. We assess how the currently observed signal could eventually limit future capabilities to search for GW relics of primordial phenomena and associated phenomenological consequences such as primordial black hole (PBH) formation. 
Given the sensitivity of PBH abundance to spectral parameters, measuring it remains a challenge for realistic signals. However, future observation could still rule out nearly subsolar mass PBHs formed through standard formation scenarios in some cases.
Future progress in constraining PBH models is expected to stem from theoretical advancements in PBH computations, which should help resolve the tension between different computational methods.
The analysis is based on and extends the Python code \href{https://github.com/Mauropieroni/fastPTA/}{\texttt{fastPTA}}.
\end{abstract}

\maketitle

\preprint{CERN-TH-2025-045} 

\hypersetup{linkcolor=bluscuro}

\section{Introduction}
\label{sec:introduction}

Pulsar timing array (PTA) experiments represent an unprecedented opportunity to investigate the spectrum of gravitational waves (GWs) in the nHz frequency range. 
Recently, various PTA collaborations reported substantial evidence of a stochastic common spectrum with an angular correlation pattern consistent with the quadrupolar nature of GWs in general relativity~\cite{Hellings:1983fr}. As reported by NANOGrav (NG)~\cite{NANOGrav:2023gor}, EPTA (in combination with InPTA)~\cite{EPTA:2023fyk}, PPTA~\cite{Reardon:2023gzh} and CPTA~\cite{Xu:2023wog}, the evidence reaches $(2\div 4) \sigma$ significance and is expected to grow in the not so distant future \cite{Siemens:2013zla,NANOGrav:2020spf,Babak:2024yhu}. Most recently, the MeerKAT PTA (MPTA) reported similar results \cite{Miles:2024seg}. 
The excess is compatible with a stochastic GW background (SGWB) characterized by a
blue-tilted spectrum of energy density $\Omega_\text{\tiny GW}\propto f^{n_T}$, 
with a best-fit value of $n_T\simeq 2$ and relatively large uncertainties.

In the astrophysical setting, this signal can be produced by the incoherent superposition of GWs emitted by a population of inspiralling supermassive black hole (SMBH) binaries \cite{EPTA:2023xxk,NANOGrav:2023hfp}. While $\Omega_\text{\tiny GW} \propto f^{2/3}$ for circular orbits in the GW-driven regime,
eccentricity and environmental effects can modify binary evolution. Additionally, the discrete number of sources and statistical fluctuation thereof in each frequency bin can lead to different predictions~\cite{Sesana:2008mz,Kocsis:2010xa,Ellis:2023owy}.

In a cosmological setting, plausible sources of nHz frequencies are extremely diverse. For example, first-order phase transitions,
cosmic strings, domain walls,
scalar-induced GWs (SIGWs),
linear tensor perturbations generated during inflation, 
among others, could contribute to the observed signal, as addressed by an impressive body of literature (see Ref.~\cite{Caprini:2018mtu} for a review). 
Inflationary models capable of amplifying scalar fluctuations are particularly promising candidates for generating an SGWB within the frequency range detectable by PTAs. 
These enhanced scalar fluctuations lead to the production of SIGWs at second-order in perturbations~\cite{Tomita:1975kj, Matarrese:1992rp,Matarrese:1993zf,Matarrese:1997ay, Acquaviva:2002ud, Mollerach:2003nq,Carbone:2004iv, Ananda:2006af, Baumann:2007zm, Domenech:2021ztg}, which could give rise to a significant SGWB.
Amplified scalar perturbations can naturally emerge from single-field inflationary models with features in the potential, such as those inducing ultra-slow-roll (USR) phases. They may also arise in multi-field inflationary scenarios or due to mechanisms like preheating and early matter-dominated periods. 

Notably, the same scalar perturbations responsible for SIGWs are linked to the potential formation of primordial black holes (PBHs)~\cite{Zeldovich:1967lct,Hawking:1971ei,Carr:1974nx,Carr:1975qj,Chapline:1975ojl,Byrnes:2025tji}. 
When SIGWs fall in the nHz frequency band, they correspond to the stellar mass range of PBHs, which can contribute to a part of the dark matter and are complementarily searched for through lensing and ground-based GW searches looking for mergers (see Ref.~\cite{Carr:2020gox} for a review).
Through the detection of SIGWs or by establishing constraints on their amplitude, PTA data has the potential to provide crucial insights into the inflationary epoch, 
potentially strengthening bounds on dark matter candidates.
On the other hand, the overproduction of dark matter in the form of PBHs in the stellar mass range implies a bound around ${\cal P}_\zeta \leq \mathcal{O}(10^{-2})$~\cite{Saito:2008jc,Bugaev:2009kq,Bugaev:2010bb,Byrnes:2018txb,Gow:2020bzo} on the scalar curvature perturbations and thus also on the strength of the SIGW in the nHz frequency band. 
For all these reasons, SIGWs are a compelling and high-priority observational target for PTAs. Thus, exploring PTA capabilities to constrain such a scenario is timely, and including theoretical information on SIGWs can help interpret current and future data.

Currently, many cosmological and astrophysical models compatible with the data predict GW spectra with similar shapes, and existing measurement uncertainties do not allow us to distinguish between them~\cite{NANOGrav:2023hvm,EPTA:2023xxk,Madge:2023dxc,Figueroa:2023zhu,Ellis:2023oxs}. 
Therefore, it is possible 
that SIGW is the dominant source, or that an astrophysical background dominates PTA observations, and SIGWs searches would need to dig below the astrophysical foreground, as we will show. We do not consider the possibility of multiple cosmological signals, although the same logic would apply in that case too. 
We will show that future datasets will improve our current understanding, showing however how systematic uncertainties on the computation of the PBH abundance remain and could limit our ability to draw stringent conclusions (see also \cite{Dandoy:2023jot,Iovino:2024tyg}). 

This paper is organized as follows. 
In Sec.~\ref{sec:2} we discuss the phenomenological parametrization we adopt of scenarios with enhanced curvature power spectra at small scales, including how to compute the spectrum of SIGW. 
In Sec.~\ref{sec:PTA_forecast} we describe the essence of PTA searches for GWs, and the construction of the PTA likelihood that we adopt to estimate current and future sensitivity to SIGWs. 
In Sec.~\ref{sec:futureBGWs} we forecast the future bound on both motivated astrophysical and SIGW signals and conclude the ability of future PTAs to constrain the spectrum of curvature perturbations. Moreover, in Sec.~\ref{sec:implications} we connect these results to the related abundance of PBH, commenting on the challenges associated with setting robust bounds due to the large model dependence and systematic uncertainties.
We conclude in Sec.~\ref{sec:conclusions} with some discussion and outlook.

\section{Second-order induced GWs}\label{sec:2}

The generation of GWs from scalar (curvature) perturbations is a direct consequence of general relativity. While scalar and tensor modes do not couple at first-order in perturbation theory, second-order tensor perturbations are sourced by first-order scalar modes. However, producing an observable SGWB requires sufficiently large scalar perturbations. Although the amplitude and shape of the curvature power spectrum are tightly constrained by cosmic microwave background (CMB) and large-scale structure (LSS) observations on scales $k \lesssim \mathrm{Mpc}^{-1}$, they remain unknown at smaller scales.
The frequency range probed by PTAs corresponds to scales $k\sim (10^6 - 10^8)\,\mathrm{Mpc}^{-1}$ which entered the Hubble sphere around the epoch of the QCD crossover~\cite{Schwarz:1997gv,Abe:2020sqb,Dandoy:2023jot,Franciolini:2023wjm}. 
Therefore, measuring SIGWs at PTA scales indirectly probes small-scale enhancements of the curvature power spectrum at such scales.

The current day abundance can be expressed as (see~\cite{Caprini:2018mtu})
\begin{equation}\label{eq:GW}
    h^2 \OGW (k) = \frac{h^2 \Omega_r }{24} \left(\frac{g_*}{g_*^0} \right) \left(\frac{g_{*s}}{g_{*s}^0}\right)^{-\frac{4}{3}} \mathcal{P}_h (k),
\end{equation}
as a function of the tensor power spectrum ${\cal P}_h(k)$, 
where $g_{*s} \equiv g_{*s} (T_k)$ and $g_* \equiv g_* (T_k)$ represent the effective entropy and energy degrees of freedom at the time of Hubble crossing of mode $k$ and at present (denoted by superscript $0$), respectively. The current radiation abundance is $h^2\Omega_r = 4.2\times 10^{-5}$, while $g_{*}^0=3.38$ and $g_{*s}^0=3.93$.  

Each mode $k$ crosses the Hubble sphere at a temperature $T_k$ given by 
\begin{equation}
    k = 1.5 \times 10^7 \, {\rm Mpc^{-1}} \left(\frac{g_*}{106.75} \right)^{\frac{1}{2}} \left(\frac{g_{*s}}{106.75}\right)^{-\frac{1}{3}} \left(\frac{T_k}{\rm GeV}\right),
\end{equation}
and corresponds to a current GW frequency 
\begin{equation}
    f \equiv k/2 \pi = 1.6 \, {\rm nHz} \left( \frac{k}{10^6 \, {\rm Mpc}^{-1}} \right).
\end{equation}
Finally, limiting ourselves to Gaussian curvature perturbations, the tensor mode power spectrum is expressed as~\cite{Kohri:2018awv,Espinosa:2018eve}
\begin{multline}
\overline{\mathcal{P}_h (\eta, k)} = 4 
\int_0^\infty \text{d}t \int_{0}^{1}\text{d} s \left [ \frac{t(2+t)(1-s^2)}{
(1-s+t)(1+s+t)
} \right ]^2 
\\
\overline{I^2 (t,s,k,\eta)} 
\times \mathcal{P}_\zeta \left(u k\right) 
\mathcal{P}_\zeta \left(v k \right) , 
\label{eq:P_h_ts}
\end{multline}
where the transfer function assuming radiation domination (RD) can be written as
\begin{align}
&\overline{I_{\text{RD}}^2(t, s, k\eta \to \infty)} =\frac{1}{2 (k \eta)^2}
I_A^2(u,v)\left[I_B^2(u,v)+I_C^2(u,v)\right] 
\label{I_RD_osc_ave_ts}
\end{align}
where
\begin{align}
I_A(u,v)&= \frac{3(u^2+v^2-3)}{4u^3v^3} \,,
\nonumber \\
I_B(u,v)&= -4uv+(u^2+v^2-3)\log\left|\frac{3-(u+v)^2}{3-(u-v)^2}\right| \,,
\nonumber \\
I_C(u,v)&= \pi(u^2+v^2-3)\Theta(u+v-\sqrt{3}) \,,
\label{IABC-functions}
\end{align}
with $u \equiv (1 + s + t)/2$, 
and $v \equiv (1 - s + t)/2$.

To simplify the analyses, we assume that the nHz SIGW background is emitted during an approximately radiation-dominated era in the early universe, ignoring EoS and sound speed variation during the QCD era~\cite{Hajkarim:2019nbx, Abe:2020sqb}, which imprints specific features on the low-frequency tail of the SGWB~\cite{Franciolini:2023wjm}.\footnote{The abundance of late-time universe SIGWs emitted during a RD phase is gauge-independent~\cite{Inomata:2019yww,DeLuca:2019ufz,Yuan:2019fwv,Domenech:2020xin}.}
Additionally, cosmic expansion may be influenced by unknown dark sector physics, potentially causing transient matter domination or kination~\cite{Ferreira:1997hj,Pallis:2005bb,Redmond:2018xty,Co:2021lkc,Gouttenoire:2021jhk,Chang:2021afa}. Such non-standard cosmologies can significantly affect SIGW and PBH production~\cite{Dalianis:2019asr,Bhattacharya:2019bvk,Bhattacharya:2020lhc,Ireland:2023avg,Bhattacharya:2023ztw,Ghoshal:2023sfa,Domenech:2019quo,Domenech:2024drm,Kumar:2024bvp}. We leave the exploration of these scenarios for future work.

\subsection{SIGW template for a lognormal spectrum }

The spectrum of SIGWs depends on the features of the curvature power spectrum ${\cal P}_{\zeta}(k)$, as described in Eq.~\eqref{eq:P_h_ts}. 
This is fully model-dependent. In the simplest scenarios, for example,  the enhanced curvature perturbations are induced by features in the potential of single-field inflation. See, e.g., Ref.~\cite{Ozsoy:2023ryl} for a review on model building. 
One could perform the analysis by adopting different approaches. In particular, one could consider an explicit model of the early universe, and compute the curvature power spectrum from first principles, varying the model parameters. On the other hand, as often done in the literature, one can adopt a template-based method, where the shape of the curvature power spectrum is assumed to follow a given functional form, inspired by an explicit model, which is controlled by phenomenological parameters (see, e.g.,~\cite{LISACosmologyWorkingGroup:2025vdz} for a comparison of these approaches in the context of SIGWs in LISA). 
We will follow the latter approach and remain agnostic on the model producing the enhancement of curvature perturbations at small scales.
Therefore, we consider a log-normal parametrization of the curvature power spectrum of the form\footnote{The choice of normalization of the amplitude $\propto \Az/\Delta$ is standard in the literature, as it makes the variance of perturbations finite in the small width limit. However, this choice inevitably introduces degeneracies between the overall $\Az$ amplitude and $\Delta$.}
\begin{equation}\label{eq:Curv Power Spectrum}
{\cal P}_{\zeta}(k)=\frac{\Az}{\sqrt{2\pi}\Delta}\text{exp}\left[-\frac{\text{log}^2(k/k_\star)}{2\Delta^2} \right].
\end{equation}
This assumption allows us to speed up the computation of the double integral in Eq.~\eqref{eq:P_h_ts} and express the spectrum of SIGWs analytically, as~\cite{Pi:2020otn} (see also~\cite{Dandoy:2023jot})
\begin{widetext}
\begin{align}\label{eq:GW broad}
h^2\Omega_{\text{GW}}(f)
\approx 
\frac{4}{5\sqrt{\pi} \Delta} 
c_g(f)
A_{\zeta}^2 \kappa ^3  
\text{exp}^{{\frac{9\Delta^2}{4}}}
\left[\left({\cal K}^2 +\frac{\Delta^2}{2}\right)
\right.
\left.\text{erfc}\left(\frac{{\cal K}+\frac{1}{2}\text{log}\frac{3}{2}}{\Delta}\right)
-
\frac{\Delta}{\sqrt{\pi}} \text{exp}\left( -\frac{\left({\cal K}+\frac{1}{2}\text{log}\frac{3}{2}\right)^2}{\Delta^2}\right) 
 \left({\cal K}-\frac{1}{2}\text{log}\frac{3}{2}\right) \right] 
\nonumber \\
+ \frac{0.0659 \kappa^2 }{\Delta^2}e^{\Delta^2}
\text{exp}
\left[-\frac{\left(\text{log}\kappa+\Delta^2-\frac{1}{2}\text{log}\frac{4}{3}\right)^2}{\Delta^2}\right]
+ \frac{1}{3}\sqrt{\frac{2}{\pi}}\frac{e^{8\Delta^2}}{\kappa^{4}\Delta}\text{exp}\left(-\frac{\text{log}^2\kappa}{2\Delta^2}\right)\text{erfc}\left( \frac{4\Delta^2-\log\kappa/4}{\sqrt{2}\Delta}\right),
\end{align}
\end{widetext}
where $\kappa  =f/f_\star$, ${\cal K} = {\rm log}\left ( \kappa\,\text{exp}\left(3\Delta^2/2\right) \right )$, and we introduced
\begin{equation}
    c_g (f)= h^2 \Omega_r \left(\frac{g_*}{g_*^0} \right) \left(\frac{g_{*s}}{g_{*s}^0}\right)^{-\frac{4}{3}}.
\end{equation}
Note that Eq.~\eqref{eq:GW broad} is valid in the broad peak 
regime of the log-normal parametrization in Eq.~\eqref{eq:Curv Power Spectrum}. 
We verified that it represents a sufficiently accurate template in the range of $\Delta$ we restrict our analyses.

\section{Forecasting future PTA constraints}\label{sec:PTA_forecast}

In this section, we outline the methods adopted to estimate future PTA sensitivity, building on the framework first introduced in Ref.~\cite{Babak:2024yhu}. 
We assume the PTA data consists of residuals of time of arrivals (TOAs) $d_I(t)$, where $I = 1, \dots, \Np$ is the pulsar index. We further assume stationarity and introduce the Fourier domain data as
\be
\tilde d_i^k  = \int_{-T_{\rm obs}/2}^{T_{\rm obs}/2} d_I(t) e^{-i 2 \pi f_k t } {\rm d }t ,
\ee
where $T_{\rm obs}$ is the observation time and $f_k = k/T_{\rm obs}$ are the Fourier basis components. 
We take the data as the sum of two zero-mean, Gaussian processes: the stochastic GW signal and the noise part, $\tilde d_I^k = \tilde s_I^k + \tilde n_I^k$. Then,  all the information is contained in the covariance matrix
\be
    C_{IJ} = C_{h, IJ} + C_{n, IJ}. 
\ee
To model the SGWB, we take it as unpolarized and isotropic, which means that the Fourier modes $\tilde h_P(f,\hat k)$ of the metric spin-2 perturbation follows\footnote{See~\cite{Depta:2024ykq} for the inclusion of anisotropies.}
\be
\langle \tilde h_P(f,\hat k) \tilde h_{P'}^*(f',\hat k') \rangle
= \frac{1}{16\pi}S_h(f)\delta(f-f')\delta_{PP'}\delta^2(\hat k,\hat k')\,,
\ee
where $P = \{+, \times\}$ denotes the two polarizations and $S_h(f)$ is the (one-sided) strain power spectral density of the SGWB, related to the SGWB energy density as
\be
    \OGW h^2 = \frac{h^2}{\rho_c}\frac{d \rho_{\rm GW}}{d \log f}
    \equiv 
    \frac{2 \pi^2 f^3}{3 H_0^2/h^2} S_h,
\ee
where $\rho_c /h^2 =  3 (H_0/h)^2 /( 8 \pi G)$ is the Universe critical energy density and $H_0 /h = 1/(9.78 \,{\rm Gyr})$ is the Hubble parameter today.

Under the above assumptions, we can show that (see~\cite{Babak:2024yhu} for the details) the time delay covariance matrix is given by
\be
\langle \widetilde{ \Delta t}_I \widetilde{ \Delta t}_J^{*} \rangle 
\equiv \frac{1}{2} \delta(f-f') C_{h,IJ}
=  \frac{1}{2} \delta(f-f')  R_{IJ} S_h(f) ,
\label{e:CIJ}
\ee
where we have defined
\be
R_{IJ} = 
\chi_{IJ} \cdot  {\cal R}(f)  
\left [
{\cal T}_I(f)
{\cal T}_J(f)
{T_{IJ} }/{T_\text{obs} }
\right]^{1/2} .
\ee
The first geometrical factor $\chi_{IJ}$ is the well-known Hellings-Downs correlation pattern as a function of the angular separation between any pair of pulsars \cite{Hellings:1983fr}. The time-dependent part accounts for the pulsar's transmission function $\mathcal{T}(f)$, which models the low-frequency suppression of the extracted signal~\cite{Hazboun:2019vhv}, and the individual observation time $T_I$. The off-diagonal components are limited by the effective overlapping time between pulsars $I$ and $J$, defined as $T_{IJ}= \min[T_I, T_J]$. 
We have further defined the sky-averaged quadratic response function $\mathcal{R}(f) \equiv 1/12\pi^2 f^2$. 

We assume the noise contribution to the covariance matrix to be dominated by the diagonal terms, i.e., to be uncorrelated between pulsars. We thus approximate $C_{n, IJ} = \delta_{IJ}P_{n,I}$. 
We model the noise budget as a white noise (WN) with the addition of a red noise (RN) component, along with the temporal variations in dispersion measurements (DM) and scattering variations (SV)
\be
P_{n,I}
= P_I^{\rm WN} 
+ A^{\rm RN}_I  \left (\frac{f}{f_r} \right )^{\gamma_I^{\rm RN}}
+ P_I^{\rm DM} + P_I^{\rm SV},
\ee
where we arbitrarily fix the reference frequency at $f_r = f_{\rm yr}$ and $\gamma_I^{\rm RN}<0$.
See~\cite{Babak:2024yhu} 
for more details 
on the noise modeling adopted in \href{https://github.com/Mauropieroni/fastPTA/}{\texttt{fastPTA}},
along with its reliability in recovering the performance of current PTA datasets.

In what follows, we will simulate future PTA datasets by sampling random realizations of the noise components of the pulsars from the distribution of the {\it observed} properties of those contained in the current EPTA catalog. Also, we will not include noise parameters in the inference, but we will fix them at the best fit. While this may approximate some of the possible correlations between signal and noise, this choice was shown to have a negligible impact on the forecasts, as validated in~\cite{Babak:2024yhu}.
Also, as we extrapolate to larger datasets, the observations enter the signal-dominated regime in the most relevant frequency range, which further justifies these approximations.

The log-likelihood of Gaussian and zero-mean data $\tilde{d}^k$ entirely described by their variance $C_{IJ}(f_k, \theta)$ is given by~\cite{Bond:1998zw, Contaldi:2020rht}
\begin{align}
- \ln \mathcal{L}
(\tilde{d} \vert \theta )
= {\rm const.}
& +
\sum_{k,IJ} \bigg[
 \ln \left [ C_{IJ}(f_k,\theta)\right ] 
 \nonumber \\
&+ \tilde{d}_I^k C_{IJ}^{-1} (f_k,\theta) \tilde{d}_J^{k*} / T_{\rm obs}
\bigg], 
\label{eq:likelihood}
\end{align}
also known as Whittle's likelihood, which represents the probability of the data $\tilde{d}$ given the model parameters $\theta$. 
This is the main ingredient for the fast forecasting technique introduced in~\cite{Babak:2024yhu}, which entails evaluating the Fisher Information Matrix (FIM). The FIM is defined as
\begin{equation}
\label{eq:FIM_definition}
F_{\alpha \beta} \equiv - \left. \frac{\partial^2 \ln \mathcal{L} }{ \partial \theta^\alpha \partial \theta^\beta } \right|_{\theta =  \theta_0} = \sum_k \textrm{Tr} \left[ C^{-1}  \frac{\partial C}{\partial \theta^\alpha} C^{-1} \frac{\partial C}{\partial \theta^\beta} \right]_{\theta =  \theta_0} \; ,
\end{equation}
where $\theta_0$ is the maximum likelihood estimator of the model parameters, obtained by imposing $(\partial \ln \mathcal{L}/\partial\theta^{\alpha})\vert_{\theta_0} = 0$. For the likelihood in Eq.~\eqref{eq:likelihood}, the FIM reads
\begin{equation}
    F_{\alpha \beta} 
    \equiv 
 \sum_{f_k}
 C^{-1}_{IJ} C^{-1}_{KL} 
\frac{\partial (R_{JK} S_h)}{\partial \theta^\alpha}
\frac{\partial (R_{LI} S_h)}{\partial \theta^\beta} .
\end{equation}
The covariance matrix is then computed by inverting the FIM, and the uncertainties on the model parameters are easily obtained as $\sigma_\alpha \equiv \sqrt{F_{\alpha \alpha}^{-1}}$. This means that estimating the uncertainties boils down to inverting the FIM, with dramatically reduced computational times compared to a full-time-domain Bayesian analysis. 

The FIM is considered to be a reliable estimator of the uncertainties in the limit of large signal-to-noise ratio, for which, typically, the posterior distribution can effectively be approximated by a Gaussian distribution. When this condition is not satisfied, we will compare the FIM results with full-fledged Bayesian parameter estimation methods. To this aim, we simulate frequency-domain data $\tilde{d}^k\equiv\tilde{d}(f_k)$, where $k$ runs over the frequencies within detector sensitivity. We then generate zero-mean Gaussian realizations of signal and noise components, whose variance is defined by their spectral densities. The parameter estimation is performed via Markov Chain Monte Carlo (MCMC) using \texttt{emcee}~\cite{Foreman-Mackey:2012any}.

It is also instructive to define an effective sensitivity, which visually captures the reach of a given PTA dataset. 
Following Ref.~\cite{Babak:2024yhu}, we define (see also~\cite{Hazboun:2019vhv})
\begin{equation}\label{eq:effsens}
 S_{\rm eff}(f) = \left (C^{-1}_{IJ} C^{-1}_{KL} 
R_{JK} R_{LI}
\right ) ^{-1/2},
\end{equation}
in such a way that the SNR computation takes the familiar form ${\rm SNR}^2 =   \sum_k  \; (S_h/S_\text{eff})^2 = T_\text{obs} \int \textrm{d} f \; (S_h/S_\text{eff})^2 $.

\begin{figure*}[t!]\centering
\includegraphics[width=0.5\textwidth]{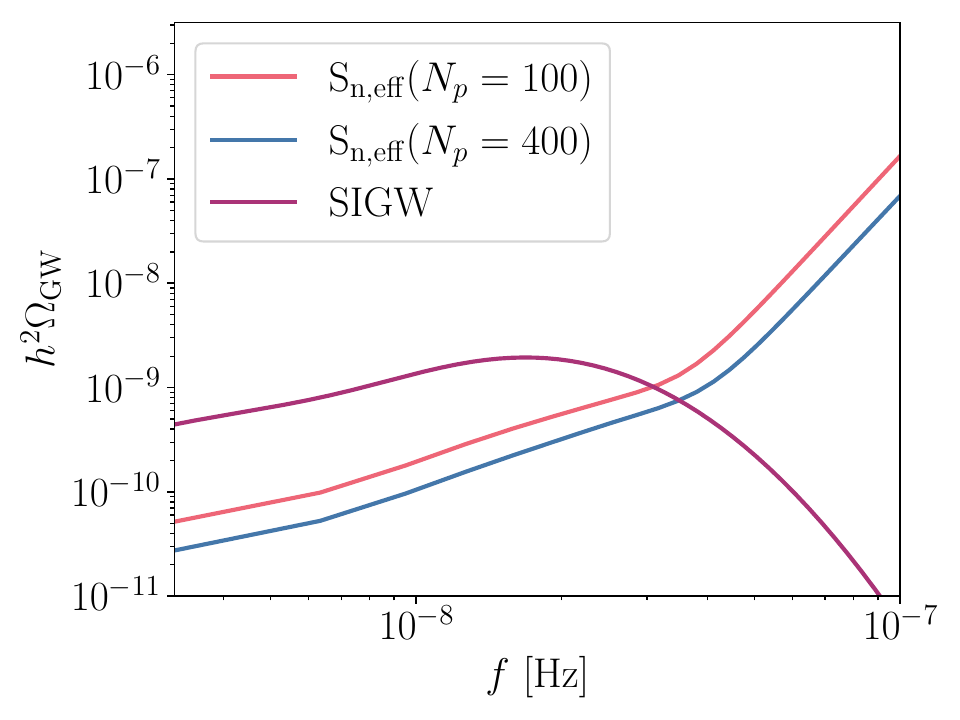}
\includegraphics[width=0.49\textwidth, trim = 0cm 0cm 0cm 0cm]{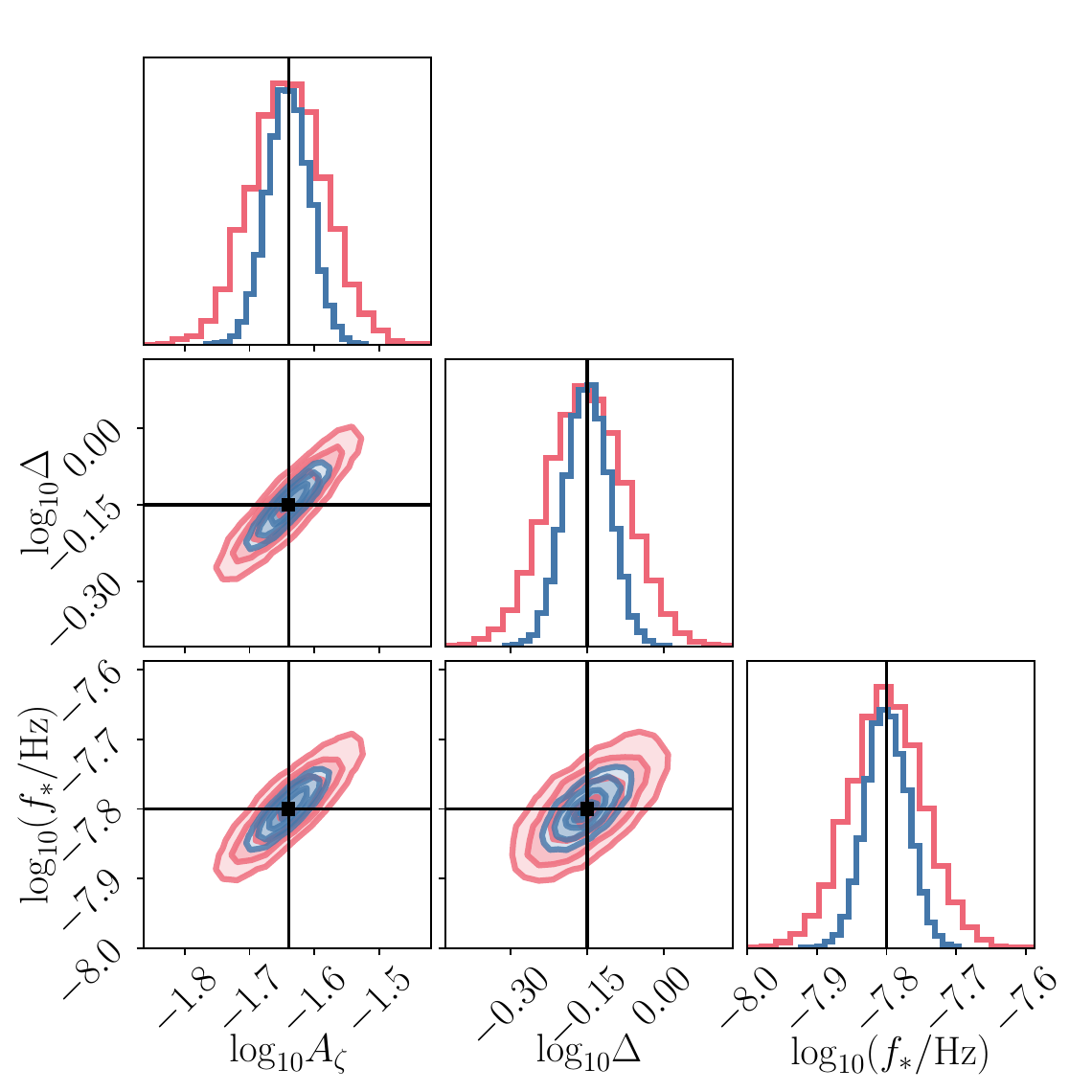}
\caption{
{\it Left panel:}
SGWB abundance for the SIGW-only scenario with 
${\rm log}_{10} \Az = -1.64$, 
$\log_{10}\Delta = -0.15$ 
and 
${\rm log}_{10} (f_*/{\rm Hz}) = -7.8 $.
With red and blue lines we indicate the effective sensitivity for a future dataset 
with $\Np =100$ and $\Np =400$ pulsar with SKA-like noise budget and $T_{\rm obs} = 10$ yr.
{\it Right panel:}
Corner plot reporting the corresponding Fisher matrix uncertainty estimates.
}
\label{fig:SIGW_only}
\end{figure*}

\section{Future bounds on SIGWs}\label{sec:futureBGWs}

In this section, we forecast the future reach of PTA datasets to constrain SIGWs and, indirectly, the primordial curvature power spectrum. 
We do so employing two working assumptions:
\begin{itemize}[leftmargin=2.5mm]

\item In the first stage (Sec.~\ref{subsec:SIGW}), we assume that the current GW excess observed in the PTA dataset is due to a cosmological SIGW. Under this hypothesis, we show how the currently large uncertainties on the amplitude of the curvature perturbation will shrink significantly with longer observation time and larger datasets.
The fact that we are observing this signal mostly in its low-frequency tail does not allow us to completely break the degeneracy between peak frequency and amplitude, leaving large uncertainties on the total amplitude of curvature perturbations.

\item In a second stage (Sec.~\ref{subsec:astro+SIGW}), we assume that the current PTA signal is dominated by an astrophysical SGWB sourced by SMBH binaries, which we model with a power law (PL), while SIGWs are only a subdominant contribution. 
As in the first case, uncertainties will shrink significantly with larger datasets and the possibility of resolving the signal peak below the dominant PL results in potentially stronger constraints on the overall SIGW amplitude. 

\end{itemize}

\subsection{SIGWs as the dominant signal}\label{subsec:SIGW}

Under the assumption of a dominant SIGW signal in the nHz frequency range, our model parameters are 
\begin{equation}\label{eq:SIGWonlyparams}
    \theta = \{\log_{10}A_{\zeta}, \log_{10}\Delta, \log_{10}(f_*/\mathrm{Hz})\}.
\end{equation} 
We simulate the observation of an SGWB dominated by SIGWs with 
\begin{align}
    \log_{10}\Az &=-1.64, 
    \nonumber \\
    \log_{10}\Delta &= -0.15,
    \nonumber \\
    \log_{10}(f_*/{\rm Hz}) &= -7.8.
    \label{eq:SIGWonlyinj}
\end{align}
We choose this set of parameters as they are compatible with the current PTA observations (NG15, EPTA+InPTA, PPTA) within $3 \sigma$ \cite{Figueroa:2023zhu,Ellis:2023oxs} while complying with the PBH overproduction bounds assuming small non-Gaussianites, see below~\cite{Franciolini:2023pbf} for details. We will come back to the PBH abundance computation in~\cref{sec:implications}. 
We checked that assuming $\Np = 68$ and current noise configurations, the expected uncertainties on the SIGW noise parameters match current constraints based on available PTA data (see, e.g.,~\cite{Franciolini:2023pbf}).

In Fig.~\ref{fig:SIGW_only}, we report the forecasts obtained with a dataset assuming SKA-like noise containing either $\Np =100$ or $\Np =400$ pulsars observed for $T_{\rm obs} = 10$ yr.
In the left panel, we show the effective sensitivity defined in Eq.~\eqref{eq:effsens}, for both cases, alongside the SIGW signal with parameters in Eq.~\eqref{eq:SIGWonlyinj}.
In the right panel, we report the posterior distribution for the model parameters, inferred using the FIM approach.
We observe a striking correlation between the amplitude $A_{\zeta}$ and the peak frequency $f_*$. 
This tight correlation arises because most of the constraining power comes from the observation of the low-frequency tail of the signal, where the signal is well approximated by a power-law, while the signal peak is more loosely constrained. 
In such a regime, there is a leftover degeneracy between the amplitude and the pivot scale, so that larger pivot scales can be compensated by increasing the amplitude of the signal accordingly. Similar correlations, albeit slightly more pronounced, are found with the width $\Delta$, due to the normalization condition in Eq.~\eqref{eq:Curv Power Spectrum}.
Overall, we conclude that given also the large degeneracies observed, the bounds on the model parameters remain rather weak. In terms of reconstructed $\OGW$, we expect it to be best constrained around $f\sim 10^{-8}$ Hz, with degeneracies maintaining fixed the best-observed region of the GW spectrum. 

\begin{figure}[t!]\centering
\includegraphics[width=0.5\textwidth]{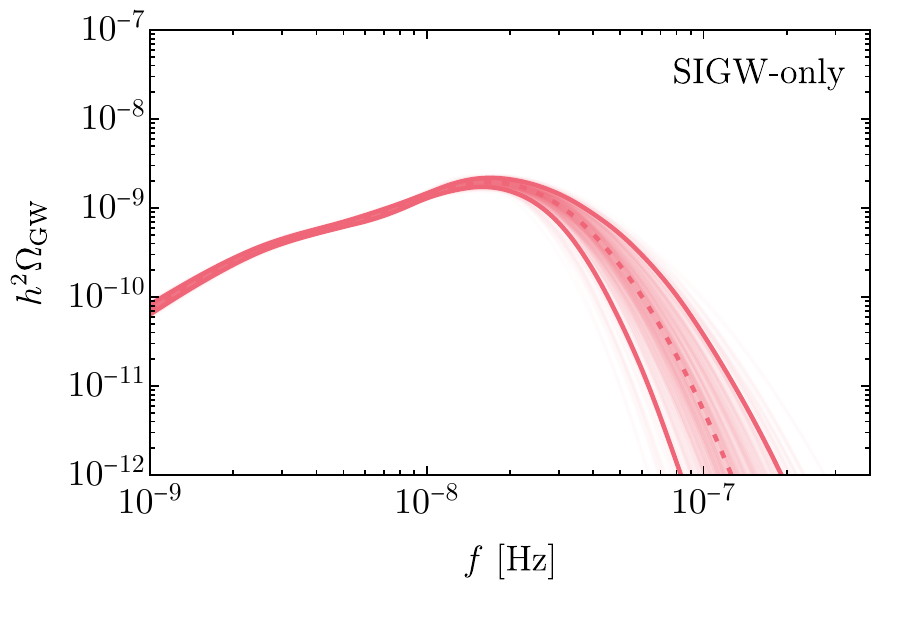}
\caption{
Posterior predictive distribution of the SIGW measured by a simulated SKA-like dataset with $\Np = 400$, as shown in Fig.~\ref{fig:SIGW_only}.
Light red lines correspond to spectra obtained by random sampling points from the posterior distribution of model parameters. The red solid line indicates the 90\% C.L., while the red dashed line is the injected $\OGW$.}
\label{fig:SIGW_only_PP}
\end{figure}

In Fig.~\ref{fig:SIGW_only_PP}, we show the posterior predictive distribution (PPD) of $\OGW$ for the case with $\Np = 400$. We define the PPD as 
\begin{equation}
p({\OGW} | d) 
= \int p({\OGW }| \theta) 
p(\theta | d) 
{\rm d}\theta
\end{equation}
where we introduced  \( p(\theta | d) \) as the posterior distribution of the model parameters \( \theta \) 
given the simulated data \( d \), while 
 \( p({\OGW} | \theta) \) is the likelihood of new signal \( \OGW \) given a set of parameters \( \theta \).
The integral marginalizes over all possible parameter values \( \theta \), weighted by their posterior probabilities.
We obtain an upper bound on the signal amplitude at all scales, as the signal is observed with sufficient SNR across the whole range of frequencies.
The high frequency tail is less constrained, as it fully depends on the width of the spectrum having $\Omega_{\rm GW} (f\gg f_*)\propto {\cal P}_\zeta^2 (f\gg f_*)$. 
Narrow spectra have a characteristic double peak feature (see e.g. discussion in \cite{Ananda:2006af,LISACosmologyWorkingGroup:2024hsc}) that leads to the observed shape, a feature which is sufficiently well reconstructed.
The reconstructed signal in the low-frequency tail is bounded to be very close to the injection, even extrapolating below the observed frequencies. This is because the causality tail does not allow for spectra narrower than $\OGW\sim f^3$~\cite{Caprini:2009fx,Cai:2019cdl,Allahverdi:2020bys,Hook:2020phx} below the most constrained region of the spectrum.

We provide fitting formulas for the precision achieved as a function of the observation time and number of pulsars. The relative uncertainties scale as 
\begin{equation}
\begin{aligned}    \frac{\sigma_{A_{\zeta}}}{A_{\zeta}} =  
    16\% 
    \left (\frac{\Np}{70} \right )^{-1/2}; \\
    \frac{\sigma_\Delta}{\Delta} =  
    21\% 
    \left (\frac{\Np}{70} \right )^{-1/2}; \\
    \frac{\sigma_{f_*}}{f_*} =  
    14\% 
    \left (\frac{\Np}{70} \right )^{-1/2}.
\end{aligned}
\end{equation}
In the scaling considered above, we do not include the fact that, as a larger and larger set of pulsars becomes available, the SNR of the signal would become larger at higher frequencies, enhancing the number of effective frequencies, thus further reducing the uncertainties in the high-frequency portion of the signal. 

Overall, we conclude that given the relatively low amplitude of the SIGW injected in this scenario, $\OGW$ would remain relatively weakly constrained also with realistic future datasets. To have tighter bounds, one would need signals with smaller $f_*$ for which the peak appears in the observable band, while still complying with the maximum ${\cal P}_\zeta$ imposed by the PBH abundance upper bound. 
Having the peak more clearly in the band would also reduce degeneracies.
We will come back to the implication of such constraints for the production of PBHs.
\color{black}

\subsection{SIGWs as a subdominant signal}\label{subsec:astro+SIGW}

In this section, we proceed with a complementary approach and assume the SIGW only amounts to a small portion of the nHz SGWB, which is instead already dominated by an astrophysical signal originating from the mergers of SMBH binaries.

\begin{figure*}[t!]\centering
\includegraphics[width=0.5\textwidth]{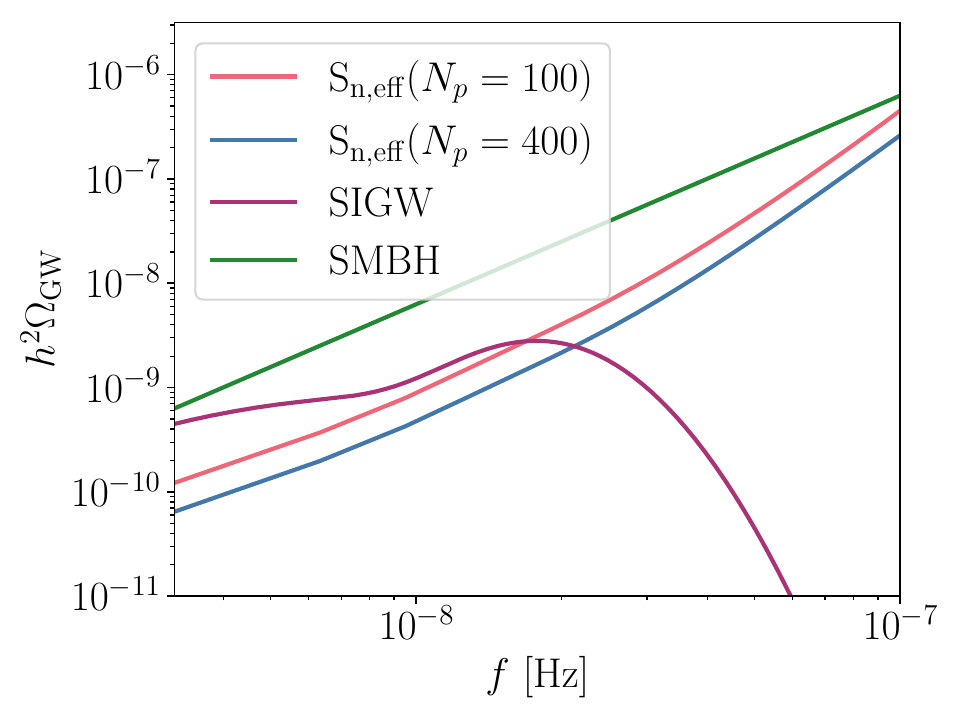}
\includegraphics[width=0.49\textwidth, trim = 0cm 0cm 0cm 0cm]{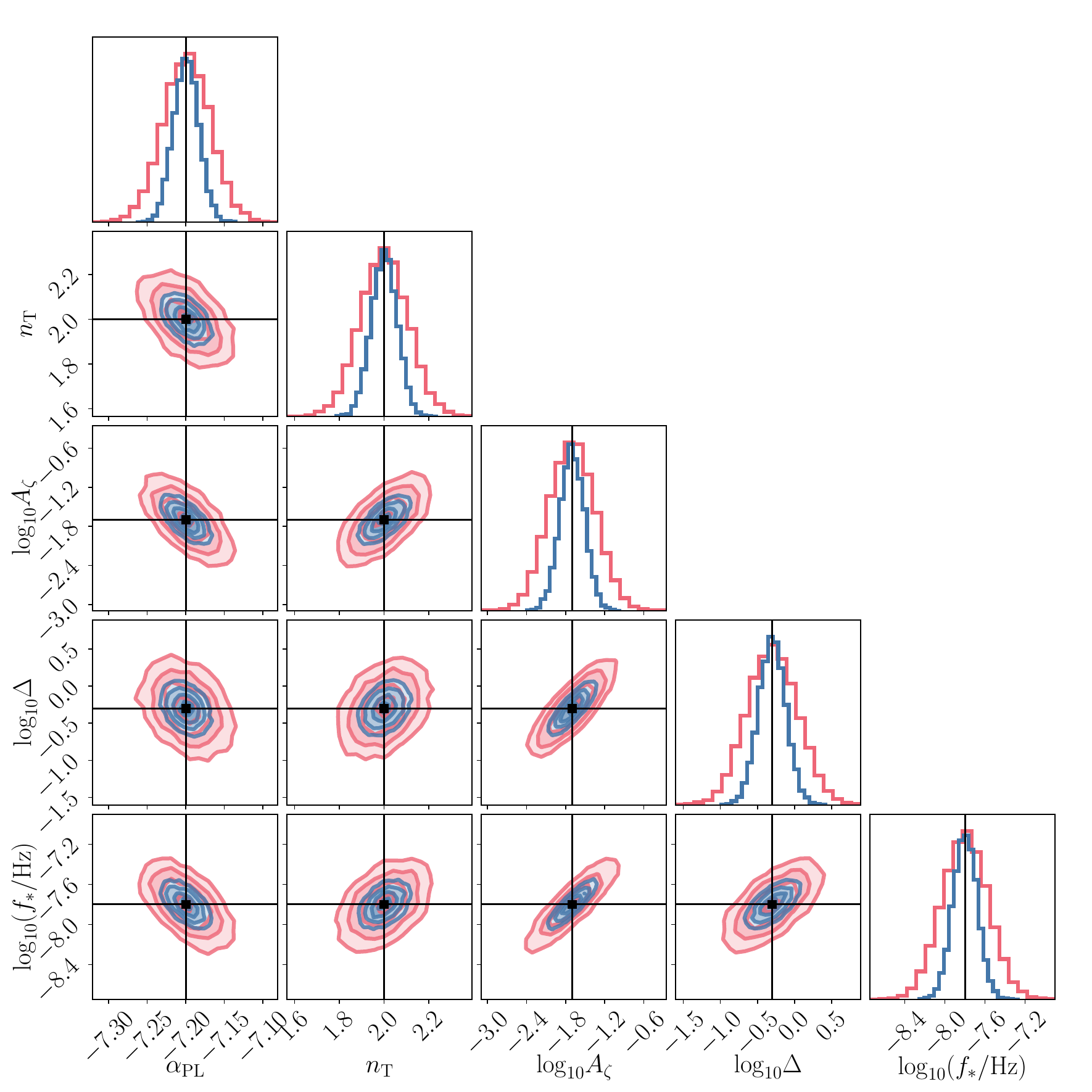}
\caption{
Same as Fig.~\ref{fig:SIGW_only}, but assuming the presence of astrophysical SGWB with 
 $\aPL = -7.2$ and $n_{T} =  2$, while 
${\rm log}_{10} \Az = -1.7$, ${\rm log}_{10} \Delta = -0.3$ and ${\rm log}_{10} (f_*/{\rm Hz}) = -7.8 $.
{\it Right panel:}
Corner plot reporting Fisher matrix uncertainty estimates for the same scenario shown in the left panel. The red (blue) posterior assumes a SKA-like with 100 (400) pulsars and $T_{\rm obs} = 10$ yr.
}
\label{fig:SIGW_plot}
\end{figure*}

\subsubsection{Modeling the astrophysical foreground}\label{sec:astromodel}

In this work, we do not attempt to model the astrophysical signal from first principles but only adopt a simple phenomenological parametrization of its spectrum. 
We, therefore, assume the astrophysical SGWB to be described by a PL compatible with the recent observations (however, see~\cite{Agazie:2024jbf}) 
\begin{equation}\label{eq:bestfitSGWB}
h^2 \OGW ^{\rm PL}(f)
= 
10^{\aPL}\left (\frac{f}{f_{\rm yr}} \right)^{n_{T}}.
\end{equation}
For definiteness, when performing the injections, the parameters are fixed to the maximum likelihood values obtained by the EPTA~\cite{EPTA:2023fyk} collaboration, which are also in good agreement with the other observations~\cite{InternationalPulsarTimingArray:2023mzf}, unless otherwise stated.
Therefore, we fix
\begin{align}\label{eq:refPL}
    \aPL &= -7.2,
    \nonumber \\
    n_{T} & =  2.
\end{align}
Here, it is important to mention that relative uncertainties on the amplitude and tilt are of the order of $60\%$ and $15\%$, respectively. 
We will explore how the result changes by varying the parameters within the $2\sigma$ range.
To make contact with other quantities used in the literature, the spectrum in Eq.~\eqref{eq:bestfitSGWB} corresponds to a strain amplitude ($h_c \equiv \sqrt{f S_h}$) at $f = f_\text{yr}$ of $A_{h_c}\simeq 10^{-14}$ and pulsar timing residuals spectral index $\gamma = 5-n_T = 3$. 

In the new physics search of the NANOGrav collaboration analysis~\cite{NG15-NP}, astrophysical population modeling was used to derive predictions for the amplitude and tilt of the SMBH SGWB spectrum, i.e., the parameters $\alpha_{\rm PL}$ and $n_T$. 
These predictions, marginalized over the SMBH population model parameters, were then used as prior in the Bayesian inference analysis, on par with the cosmological bounds on the amplitude of curvature perturbation, for example. Thus, they worked with a bivariate normal distribution for amplitude and tilt of the spectrum~\cite{NG15-NP,NANOGrav:2023hfp}. When translated in terms of $(\alpha_{\rm PL}, n_T)$,  
the mean $\mu_{\rm SMBH}$ and covariance $\sigma_{\rm SMBH}$ are given by 
\begin{equation}
\mu_{\rm SMBH} = 
\begin{pmatrix}
-10.4 \\
0.30
\end{pmatrix},
\quad
\sigma_{\rm SMBH} =
\begin{pmatrix}
1.1 & 0.0054 \\
0.0054 & 0.12
\end{pmatrix}.
\label{eq:A1}
\end{equation}
This prior distribution forces the amplitude of the SGWB to preferentially fall at smaller values than what is currently observed, thus downplaying the SMBH model in favor of new physics~\cite{NG15-NP}. In this work, we assume a more agnostic approach and use uninformative priors with distant boundaries on both parameters. 

To summarize, the total SGWB is assumed to be 
\begin{equation}
h^2\OGW = h^2\OGW^{\rm PL}+h^2\OGW^{\rm SIGW},
\end{equation}
and the model parameters in this case are
\begin{equation}
    \theta = \{\aPL, n_T, \log_{10}A_{\zeta}, \log_{10}\Delta, \log_{10}(f_*/\mathrm{Hz})\}.
\end{equation}

\subsubsection{Multi-model analysis}

We simulate a scenario where the dominant PL contribution (parameters of the PL are fixed to the reference values)
is hiding a subdominant SIGW with parameters fixed to representative values 
\begin{align}
    \log_{10}\Az &=-1.7, 
    \nonumber \\
    \log_{10}\Delta &= -0.3,
    \nonumber \\
    \log_{10}(f_*/{\rm Hz}) &= -7.8.
    \label{eq:ref_multiSIGW}
\end{align}

In Fig.~\ref{fig:SIGW_plot} (left panel), we plot the two separate contributions to the SGWB abundance and the future effective sensitivity achieved by the SKA 10yr, assuming $\Np = 100, 400$ pulsars. 
Notice the injected SIGW signal is always subdominant w.r.t. the PL in the observed frequency range. 
First of all, the presence of the dominant PL contribution significantly degrades the effective sensitivity of the simulated dataset~\cite{Babak:2024yhu}. 
However, with a sufficient number of pulsars the combined SGWB can still be measured with sufficient accuracy to allow constraining subdominant contribution with an abundance that is up to ${\cal O}(1)$ orders of magnitude below the astrophysical PL. 

We show the corresponding posterior distribution of parameters in the right panel of Fig.~\ref{fig:SIGW_plot}.
One obtains relative uncertainties of the order of ${\cal O}(20)\%$ on the curvature power spectrum
parameters, roughly indicating SIGW detection at $5 \sigma $ C.L. with more than $\Np = 400$ pulsars. 
Focusing on the SIGW parameters only, we see correlations similar to the ones observed in Fig.~\ref{fig:SIGW_only}. However, since the SIGW peak appears in the sensitivity of the simulated dataset, these correlations are less pronounced. 
There exists a negative correlation between $\aPL $ and $\Az$.
This appears since PTA datasets are sensitive to the sum of the signals, and a stronger SIGW can be compensated by a fainter astrophysical SGWB, and vice versa. Also, a similar positive correlation is observed with $n_T$, as a larger tilt removes power from the lower frequencies of relevance for the dataset we consider, due to our choice of pivot scale $f_{\rm yr}$ to normalize the PL. Finally, a negative correlation between $f_*$ and $\aPL$ appears.
This is most likely a combination of multiple trends. As a SIGW pivot at lower frequencies correlates with a narrower tilt, the steeper low-frequency SIGW behavior is correlated with a shallower PL and a larger $\aPL$.

\begin{figure}[t!]\centering
\includegraphics[width=0.5\textwidth]{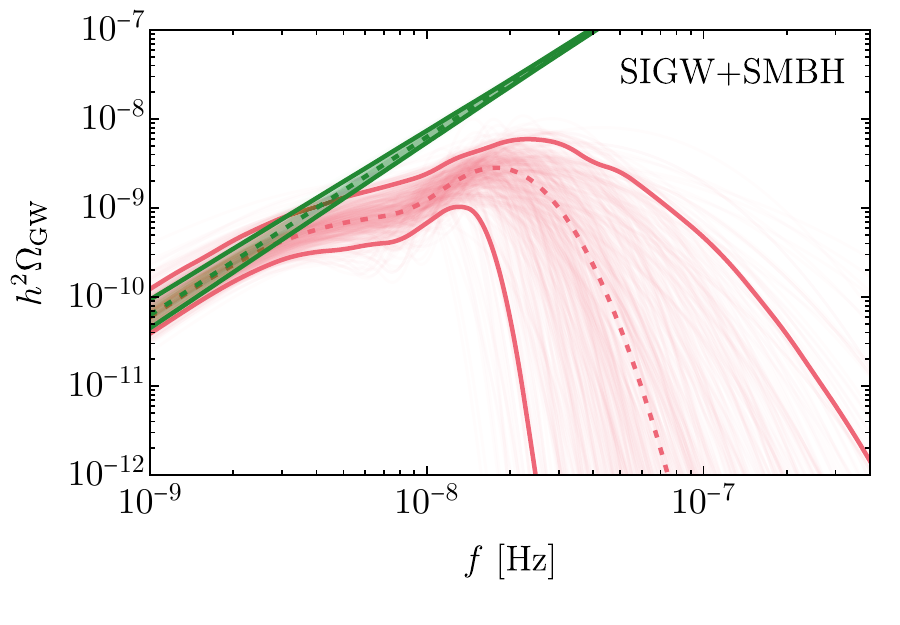}
\caption{
Same as~\ref{fig:SIGW_only_PP} for the SIGW+SMBH case. We show the posterior predictive distribution for the power law (green) and SIGW (red). We again assume a simulated SKA-like dataset with $\Np = 400$.
}\label{fig:SIGW_PL_PP}
\end{figure}

In Fig.~\ref{fig:SIGW_PL_PP}, we report the PPD obtained from the results of the analysis, assuming $\Np = 400$. 
The PL signal is very tightly constrained up to frequencies of around $f\gtrsim f\sim 10^{-7} \text{Hz}$. The subdominant SIGW signal is best constrained in the low-frequency region of the spectrum, while it becomes unbounded from below above $f\gtrsim 2\cdot 10^{-8} \text{Hz}$. As the peak of the injected signal partly appears within the sensitivity band of the dataset, the width of the spectrum is more tightly constrained, compared to the previous case (see Fig.~\ref{fig:SIGW_only_PP}). As a consequence, the uncertainty on the signal amplitude mostly rescales the signal vertically around the injected spectrum in the low-frequency range. 

While our results are subject to the assumption ${\cal P}_\zeta$ is described by a LN shape, we expect the qualitative conclusions to remain valid also for more complex shapes. It would be interesting to perform analyses with more flexible models (see e.g. \cite{LISACosmologyWorkingGroup:2025vdz}). We leave this for future work. 

It is interesting to explore how the uncertainties on the assumed subdominant SIGW background vary by modifying the assumptions. 
In the following, we explore how the uncertainties change as a function of 
\begin{itemize}[leftmargin=2.5mm]
\item [{\it i)}] the number of pulsars in the dataset; 
\item [{\it ii)}] the amplitude of the dominant astro-SGWB;
\item [{\it iii)}] the SIGW signal parameters. 
\end{itemize}

\begin{figure*}\centering
\includegraphics[width=0.245\textwidth]{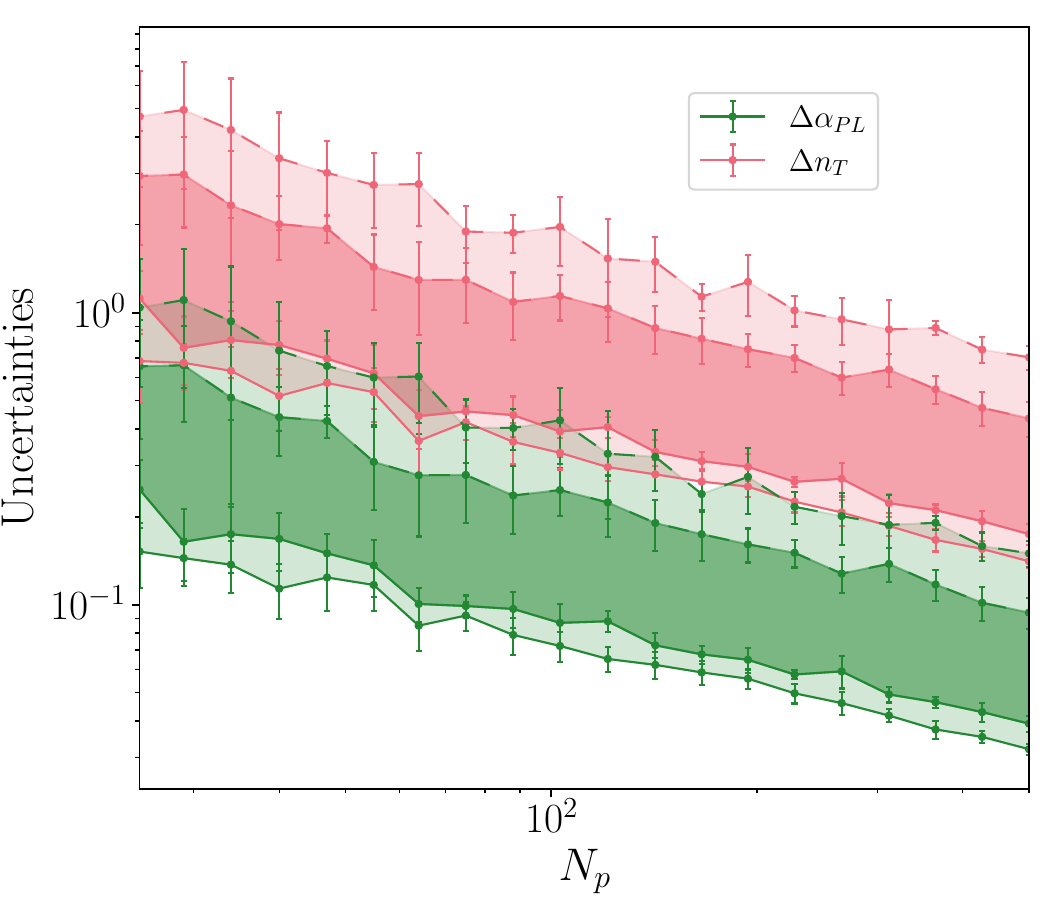}
\includegraphics[width=0.241\textwidth]{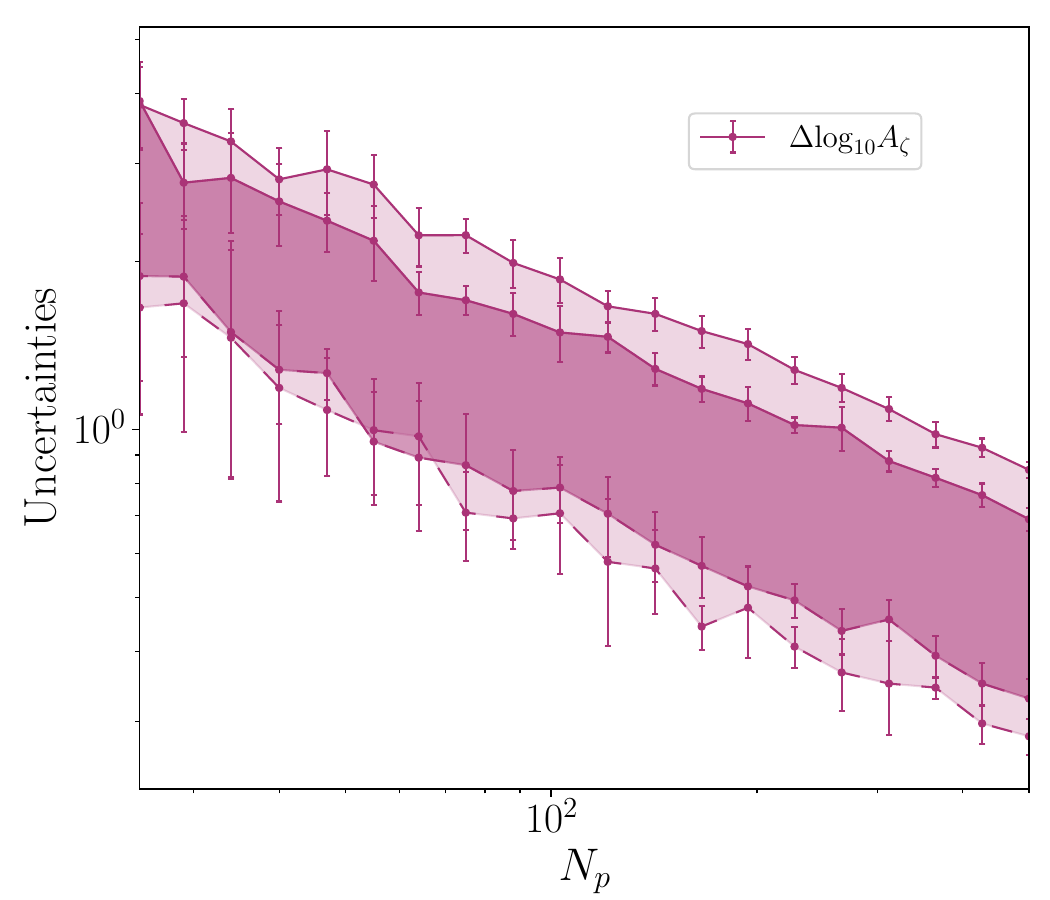}
\includegraphics[width=0.24\textwidth]{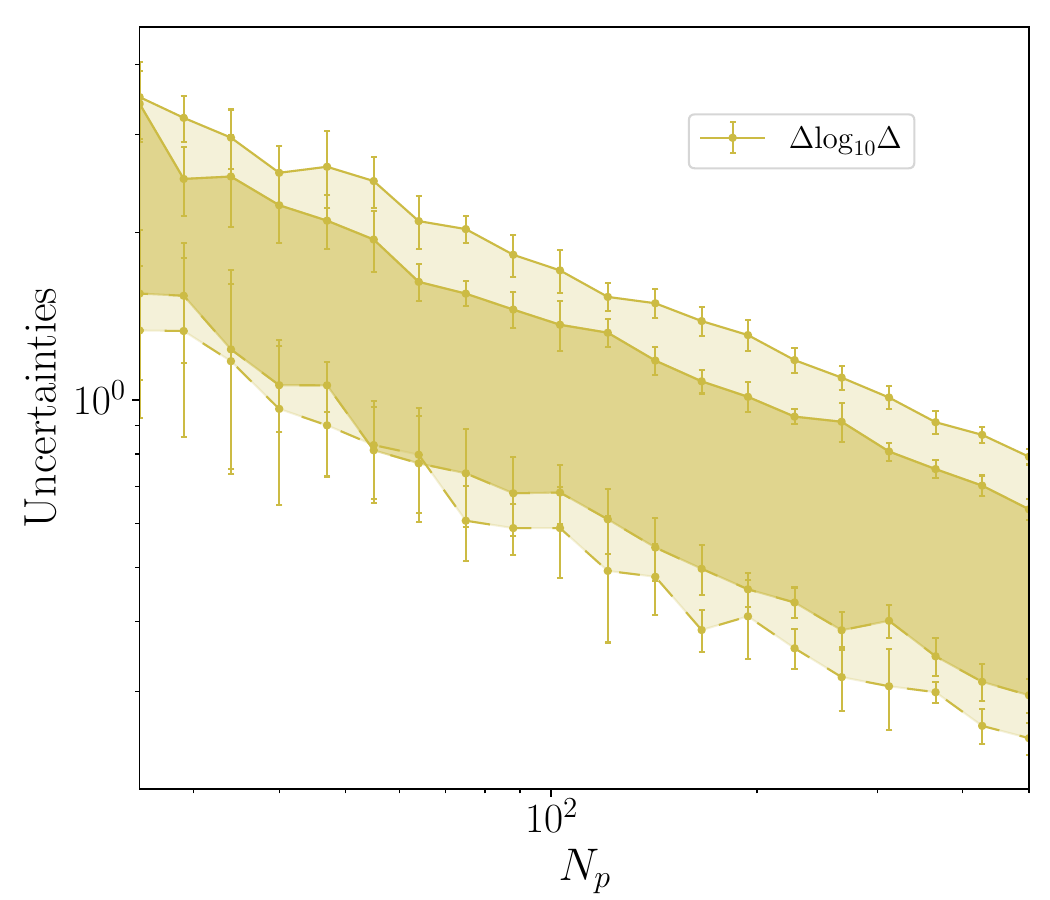}
\includegraphics[width=0.24\textwidth]{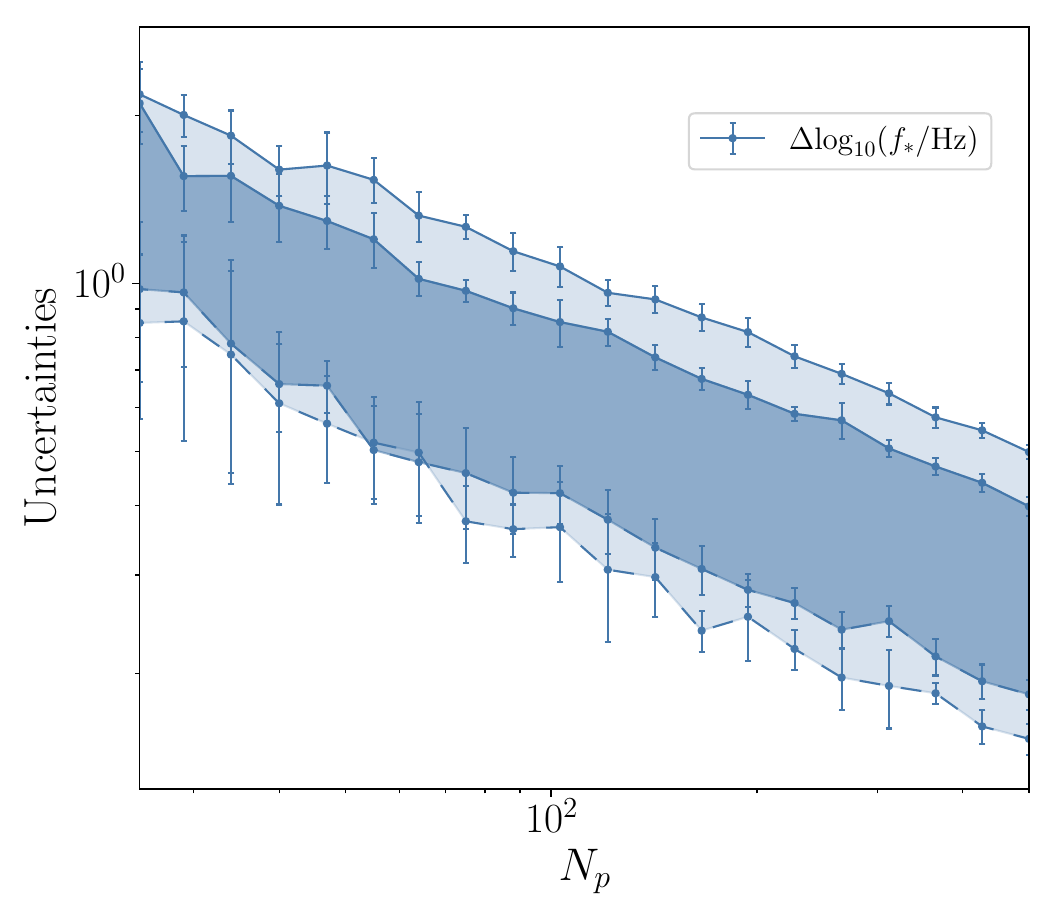}
\includegraphics[width=0.24\textwidth]{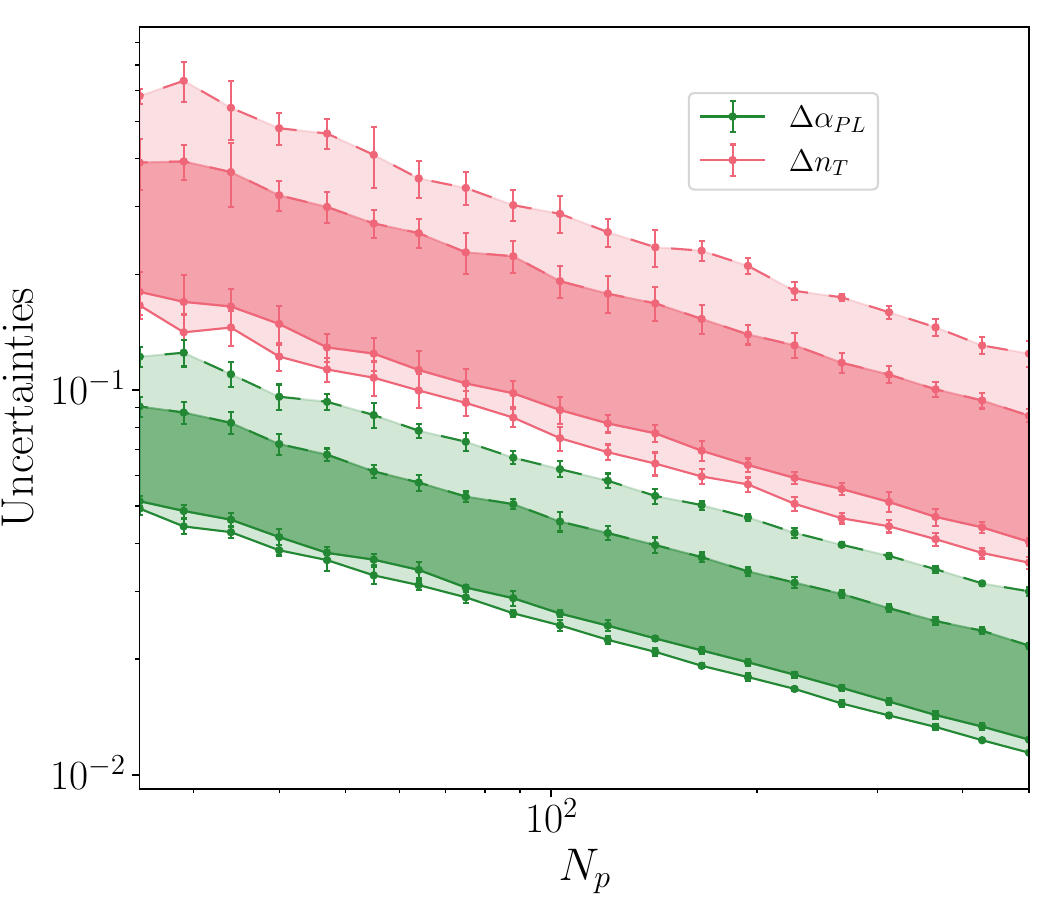}
\includegraphics[width=0.24\textwidth]{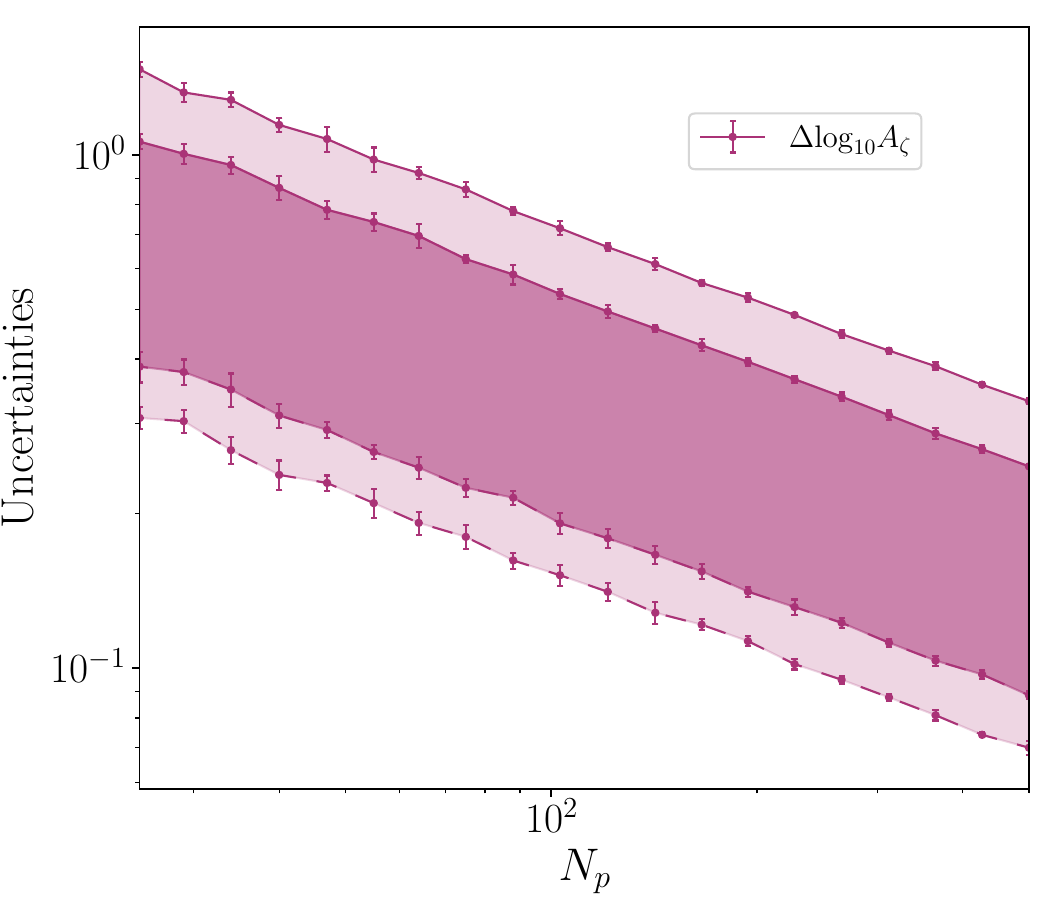}
\includegraphics[width=0.24\textwidth]{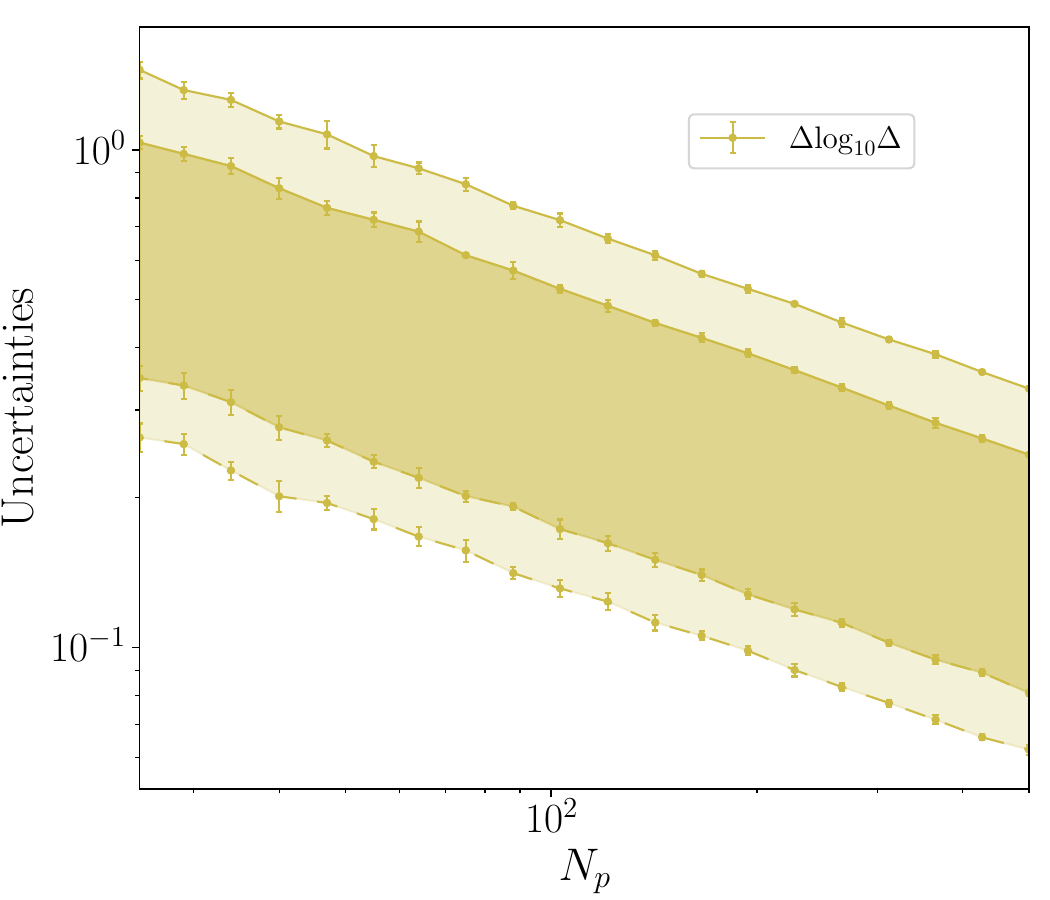}
\includegraphics[width=0.24\textwidth]{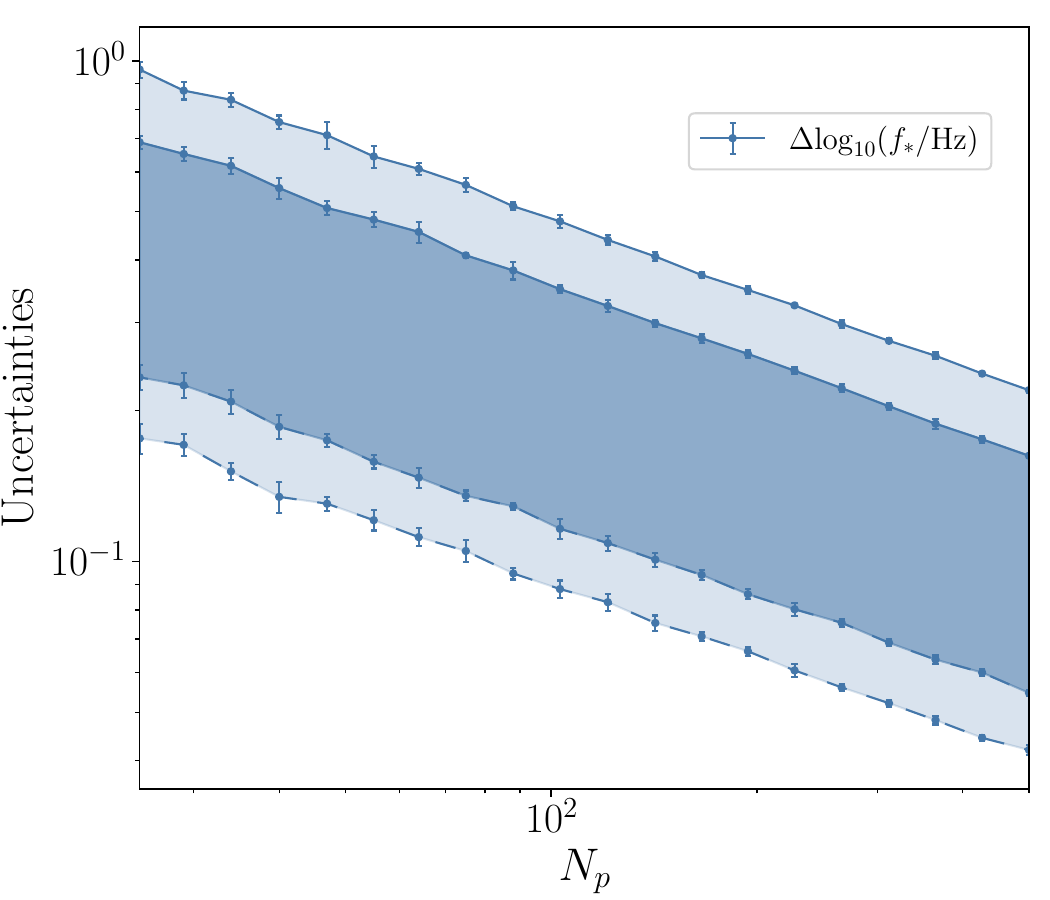}
\caption{
Scaling of the uncertainties on $\aPL$ and $n_T$ (first column), $\mathrm{log}_{10}A_{\zeta}$ (second column), $\mathrm{log}_{10}\Delta$ (third column), and $\mathrm{log}_{10}(f_*/\mathrm{Hz})$ (fourth column), as a function of the number of pulsars $\Np$, assuming a $T_{\rm obs} = 10.33$ yr.
We assume an EPTA-like ({\it top}) or SKA-like ({\it bottom}) dataset. 
The dark (light) shaded area is obtained by varying $\alpha_{\rm PL}$ within the 68\% (95\%) C.L. posterior distribution from the current EPTA+NANOGrav posterior, see~\cite{Figueroa:2023zhu}. Solid (dashed) lines correspond to larger (smaller) $\alpha_{\rm PL}$.
The bars around the band edges are obtained from the standard deviation of 10 realizations of the pulsar catalog (varying both location and noises). 
}
\label{fig:scalingNp}
\end{figure*}

\subsubsection{Scaling with the number of pulsars}

Still assuming a multi-model inference, in Fig.~\ref{fig:scalingNp} we report how the uncertainties vary as a function of the number of pulsars in the dataset $\Np$. 
We derive these uncertainties assuming the reference values of the PL parameters, as well as the SIGW parameters, as defined in Eq.~\eqref{eq:ref_multiSIGW}.
The relative uncertainties retain an overall scaling $\sim 1/\sqrt{\Np}$, as expected from the combination of $\Np$ independent datasets with spectral parameters mostly constrained by autocorrelation information (see, e.g.,~\cite{Samsing:2013kua,Babak:2024yhu}). 

We analyze the scaling behavior of the uncertainties of the following parameters: $\aPL$ and $n_T$ (first column), $\log_{10} A_{\zeta}$ (second column), $\log_{10} \Delta$ (third column), and $\log_{10} (f_*/\mathrm{Hz})$ (fourth column), as a function of the total number of pulsars, $\Np$. Our analysis assumes the observational period to be fixed to $T_{\rm obs} = 10.33$ yr.  
We consider two distinct datasets: an EPTA-like dataset (top panel) and an SKA-like dataset (bottom panel). 
The shaded regions in the plots indicate the range of variation in the uncertainties due to different choices of the PL spectral amplitude $\aPL$. 
The dark-shaded region corresponds to the 68\% C.L., i.e., $\aPL\in [-7.6, \, -7]$, while the light-shaded region extends to the 95\% C.L.,
i.e., $\aPL\in [-7.9, \, -6.8]$, 
both derived from the posterior distribution obtained using current EPTA+NANOGrav data, 
see Sec.~\ref{sec:astromodel}.  

Additionally, we differentiate between two limiting cases of $\aPL$: solid lines represent results obtained for larger values of $\aPL$, while dashed lines correspond to smaller values. As expected, in the leftmost panel, the uncertainties are larger for smaller values of $\aPL$, as it corresponds to weaker injected signals. On the other hand, in the remaining columns, the uncertainties are larger for larger values of $\aPL$, as expected for a larger dominant astrophysical foreground. 

To account for statistical variations in the pulsar population, we generate multiple realizations of the pulsar catalog. Specifically, the error bars surrounding the edges of the shaded bands reflect the standard deviation computed over 10 independent realizations, where both the spatial distribution of pulsars and their intrinsic noise properties are varied. This variability is very small for large $\Np$. Furthermore, we see that it reduces significantly when passing from EPTA to an SKA-like dataset, as the observations enter the signal-dominated regime.

We provide fitting formulas for the precision achieved as a function of the observation time and number of pulsars. The relative uncertainties of the PL parameters scale as 
\begin{align}
    \sigma_{\aPL}
    &=  [10, 24] \% 
    \left (\frac{\Np}{70} \right )^{-1/2};
    \nonumber \\
    \sigma_{n_T}
    &=
    [3.2,5.5]\% 
    \left (\frac{\Np}{70} \right )^{-1/2};
\end{align}
while the SIGW parameter uncertainties are
\begin{align}
    \frac{\sigma_{A_{\zeta}}}{A_{\zeta}} &= [52,150]\% 
    \left (\frac{\Np}{70} \right )^{-1/2};
  \nonumber 
  \\
    \frac{\sigma_\Delta}{\Delta} &=  [48,150]\% 
    \left (\frac{\Np}{70} \right )^{-1/2};
  \nonumber 
  \\
    \frac{\sigma_{f_*}}{f_*} 
    &=  [32,96] \% 
    \left (\frac{\Np}{70} \right )^{-1/2}.
\end{align}
In the above equations, we bracket the results obtained by varying $\aPL$ within the $68\%$ C.L.

As already hinted in the previous section, the relatively large uncertainties found for this injected SIGW signal, which results in a large uncertainty on the amplitude of the curvature power spectrum, imply relatively weak constraining power on the parameter space of relevance for PBH formation. 
In~\cref{sec:implications}, we will discuss the implications of such constraints for PBH production.

\begin{figure*}[t]\centering
\includegraphics[width=0.49\textwidth]{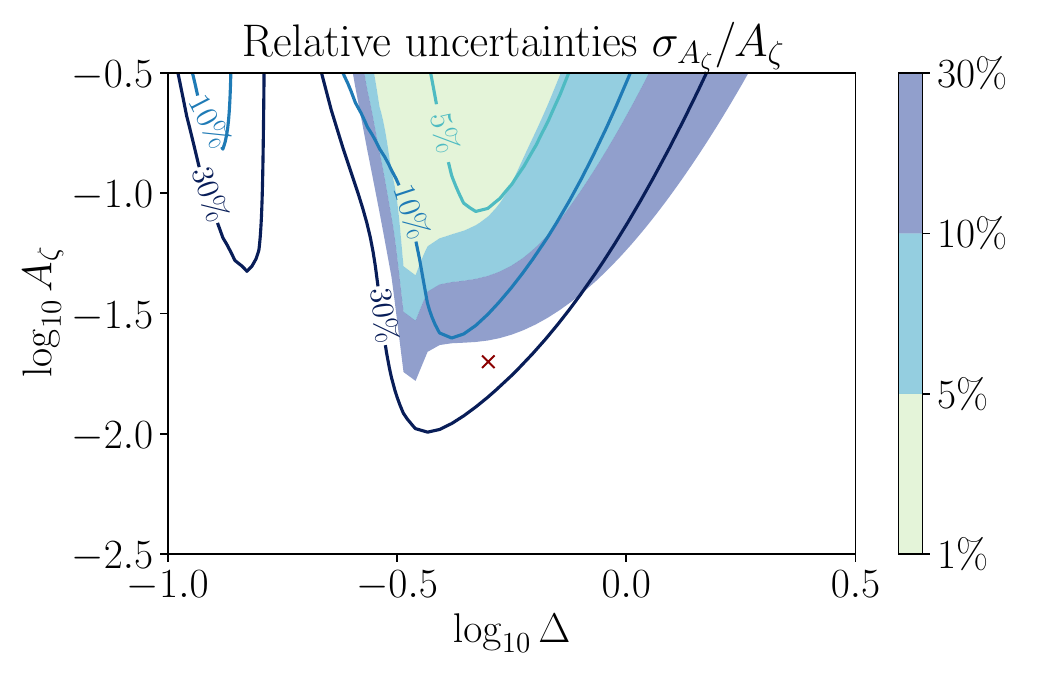}
\includegraphics[width=0.49\textwidth]{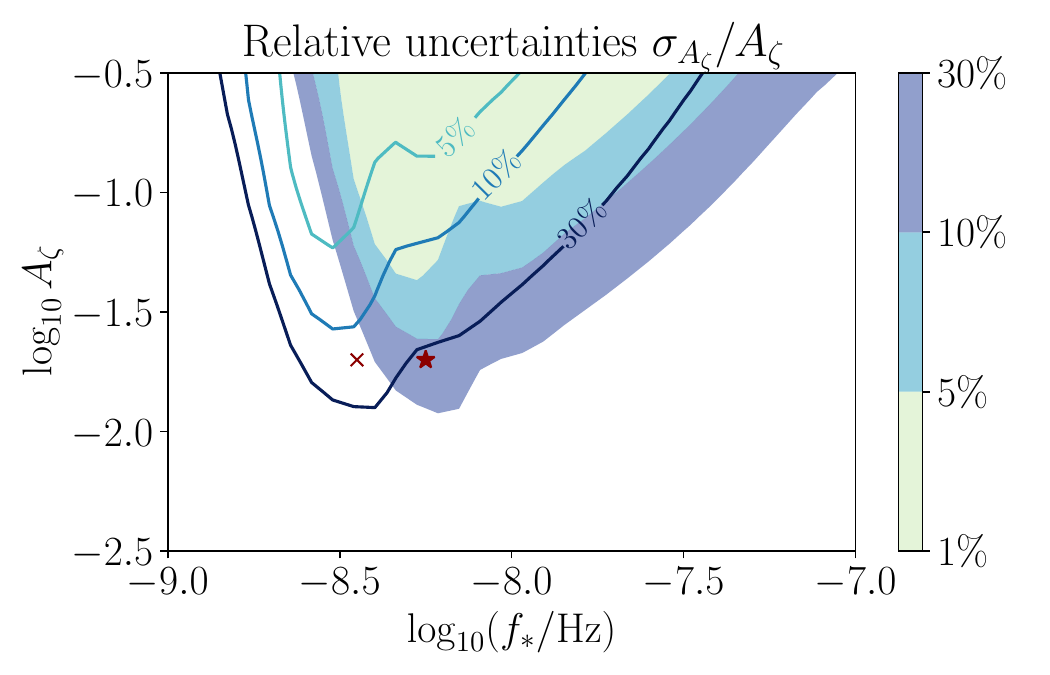}
\caption{
Fisher matrix relative uncertainties on $A_{\zeta}$ for EPTA-like configuration with $T_{\rm obs} = 16.03$ yr and $\Np = 120$ (solid lines) and SKA-like configuration with $T_{\rm obs} = 10.33$ yr and $\Np = 200$ (shaded), as a function of $\mathrm{log}_{10}\Delta$ (\textit{left panel}) and $\mathrm{log}_{10}(f_*/\mathrm{Hz})$ (\textit{right panel}).
When fixed, the parameters are set to $\log_{10}(f_*/\mathrm{Hz}) = -8.45$ (left panel) and $\Delta = 0.5$ (right panel).
The injection values for the MCMC analysis are marked with a red cross (star) for present (future) configuration. They coincide in the left panel.
}
\label{fig:Errors2D_Amplitude}
\end{figure*}

\subsubsection{Scaling with SIGW model parameters}\label{sec:scalingwithmp}

We further derive future bounds on the curvature power spectrum by varying the SIGW model parameters. 
In Fig.~\ref{fig:Errors2D_Amplitude}, we show the relative uncertainties on $\Az$ varying the injected SIGW parameters, $\Delta$ (left panel) and $f_*$ (right panel). We keep the PL parameters fixed to the reference values chosen in Eq.~\eqref{eq:refPL}.
The results for an EPTA-like dataset with $T_{\rm obs} = 16.03$ yr and $\Np = 120$, similar to the upcoming IPTA data release, are shown with a dashed line. The future SKA-like configuration with  $T_{\rm obs} = 10.33$ yr and $\Np = 200$ is shown with colored shading.

We can identify simple trends in the left plot of Fig.~\ref{fig:Errors2D_Amplitude}. We see that relative uncertainties on $\Az$ decrease both for larger amplitude and smaller $\Delta$, up to $\log_{10} \Delta\simeq -0.5$.
Due to our choice of $f_*$, narrower spectra exit from the observable window and thus correspond to larger uncertainties. The EPTA-like case, due to the longer assumed observation time, can reach a larger parameter space, as it can access slightly lower frequencies $\sim 1/T_{\rm obs}$. The secondary island is induced by the large bump in the signal obtained in the small $\Delta$ limit. 
Interestingly, in this specific setting, the uncertainties are smaller for the EPTA-like configuration. This is simply due to our choice of $f_*$.

\begin{figure*}\centering
\includegraphics[width=0.49\textwidth,
trim = 
0cm
0cm 
0cm 
1.4cm]{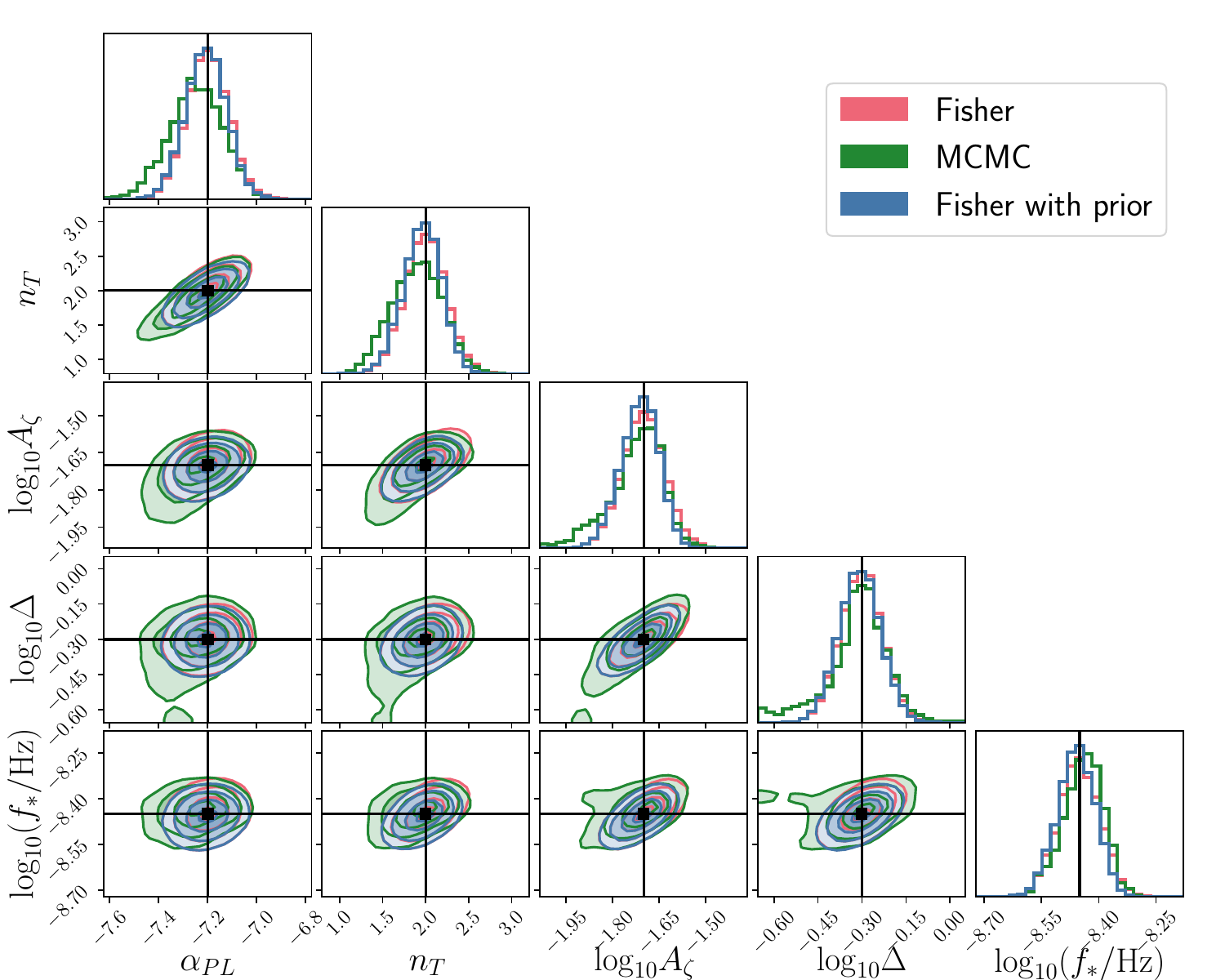}
\includegraphics[width=0.49\textwidth, trim = 0cm 0cm 0cm 1.4cm]{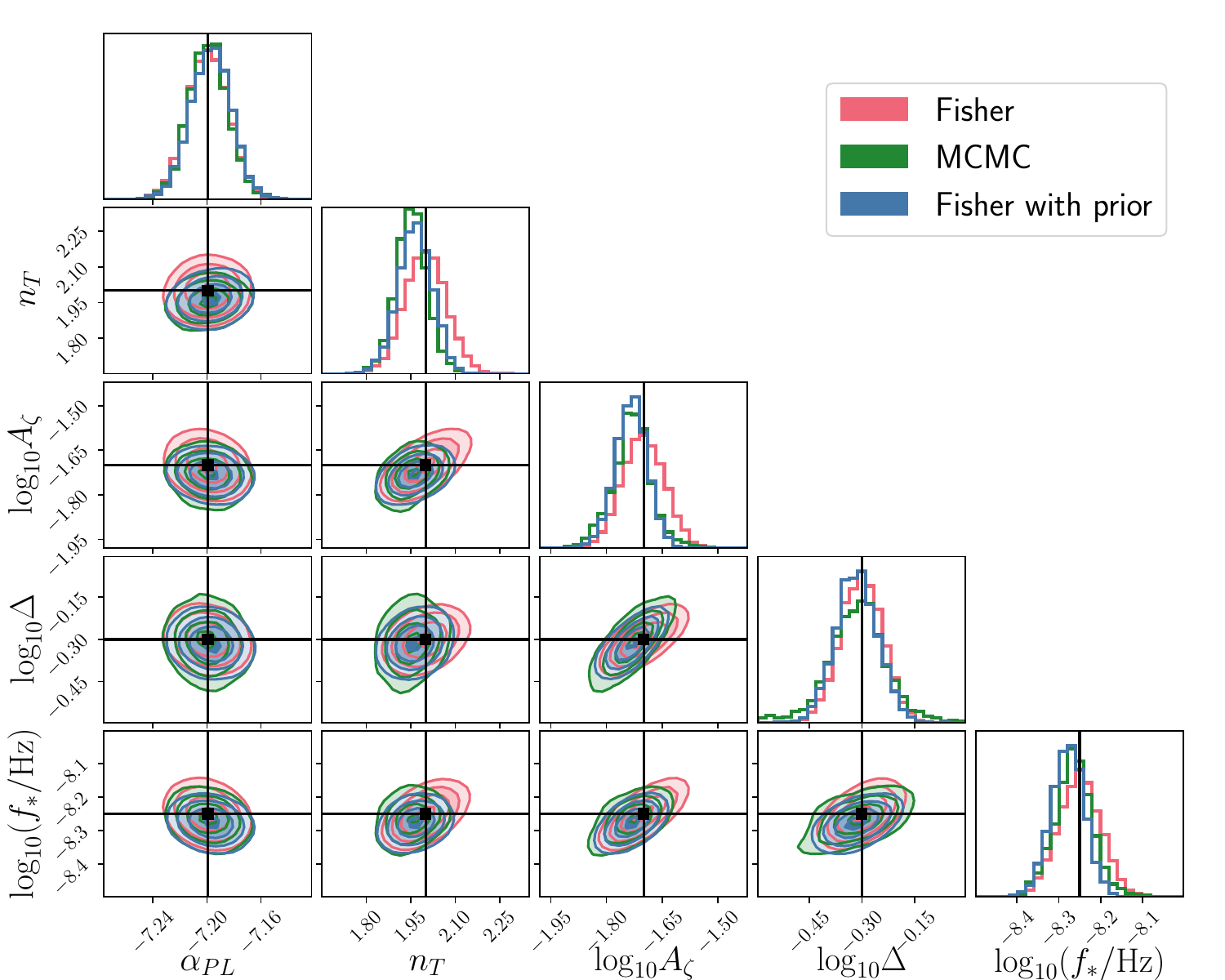}
\caption{
Comparison between 
Fisher matrix (red) and Bayesian MCMC parameter estimation (green) obtained from EPTA-like configuration with $T_{\rm obs} = 16.03$ yr and $\Np = 120$ (\textit{left panel}) and SKA-like configuration with $T_{\rm obs} = 10.33$ yr and $\Np = 200$ (\textit{right panel}).
The signal is composed of a PL as in  Eq.~\eqref{eq:refPL}, and an SIGW with $\log_{10}A_{\zeta} = -1.7, \Delta = 0.5, \log_{10}(f_*/\mathrm{Hz}) = -8.45\, (-8.25)$ for the left (right) plot. 
The Fisher matrix results excluding the points with $f_{\rm PBH}>1$ are plotted in blue. 
In the 2-d panels, $68\%, 95\%$, and $99\%$ C.I. are shown. 
The prior ranges coincide with the boundaries. 
}
\label{fig:mcmc_modelp}
\end{figure*}
\begin{figure*}[t]\centering
\includegraphics[width=0.49\textwidth,
trim = 0cm 0cm 0cm 1.cm]{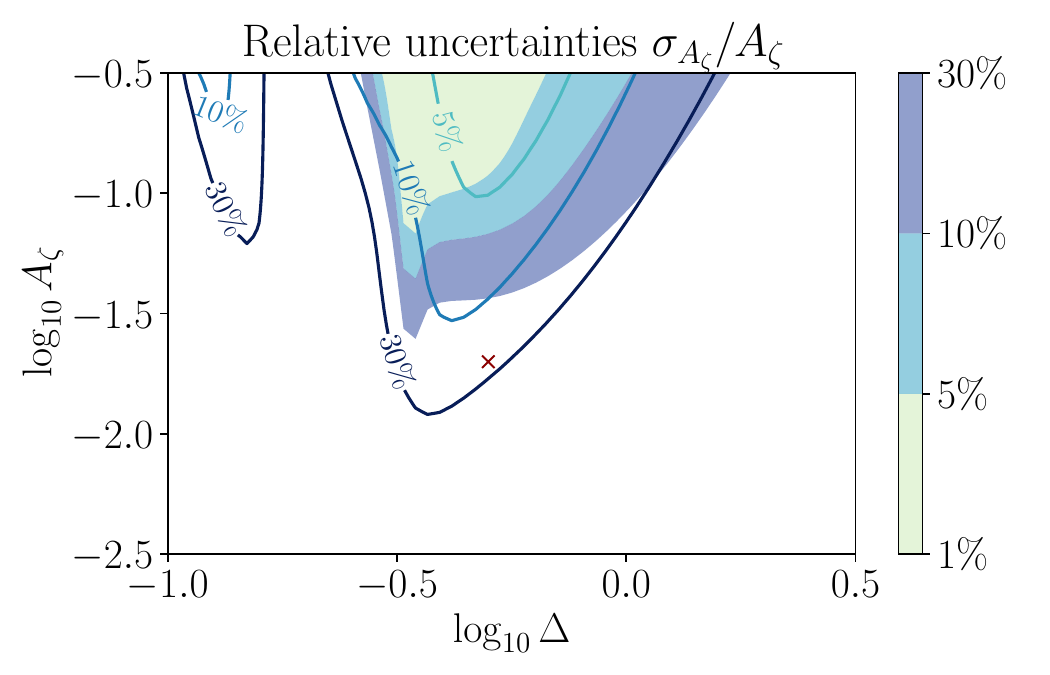}
\includegraphics[width=0.49\textwidth,trim = 0cm 0cm 0cm 1.cm]{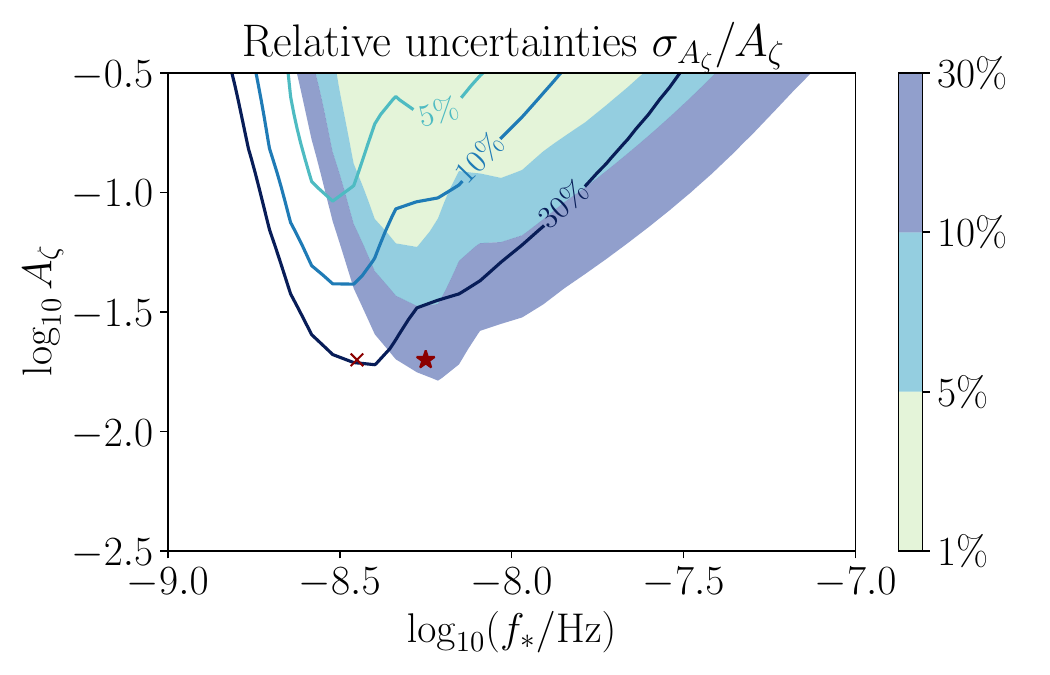}
\includegraphics[width=0.49\textwidth,trim = 0cm 0cm 0cm .4cm]{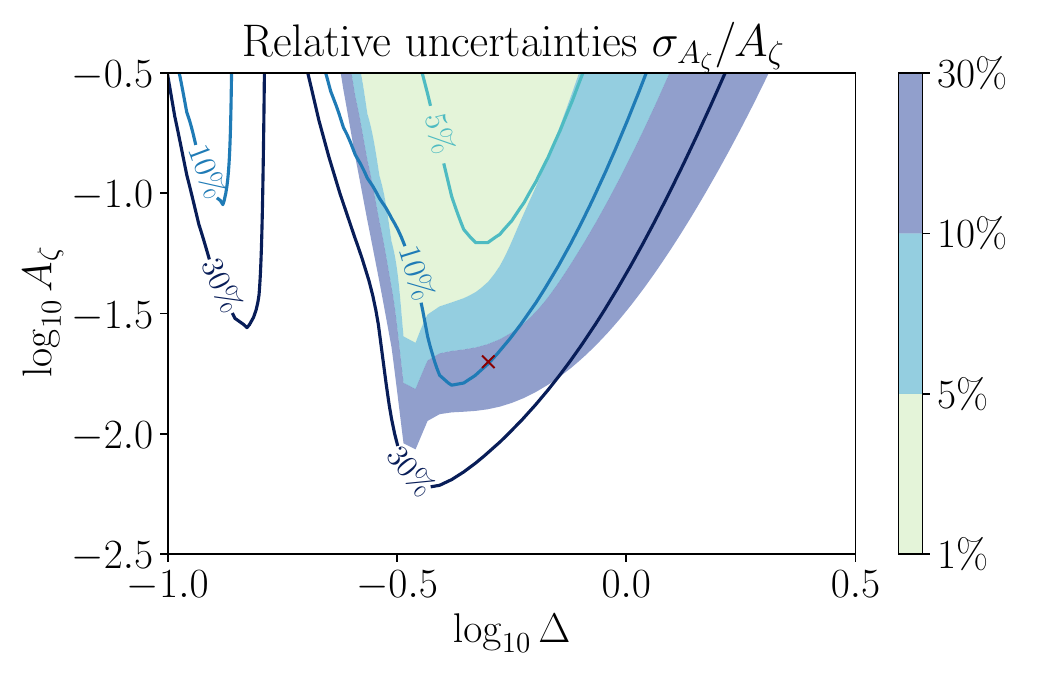}
\includegraphics[width=0.49\textwidth,trim = 0cm 0cm 0cm .4cm]{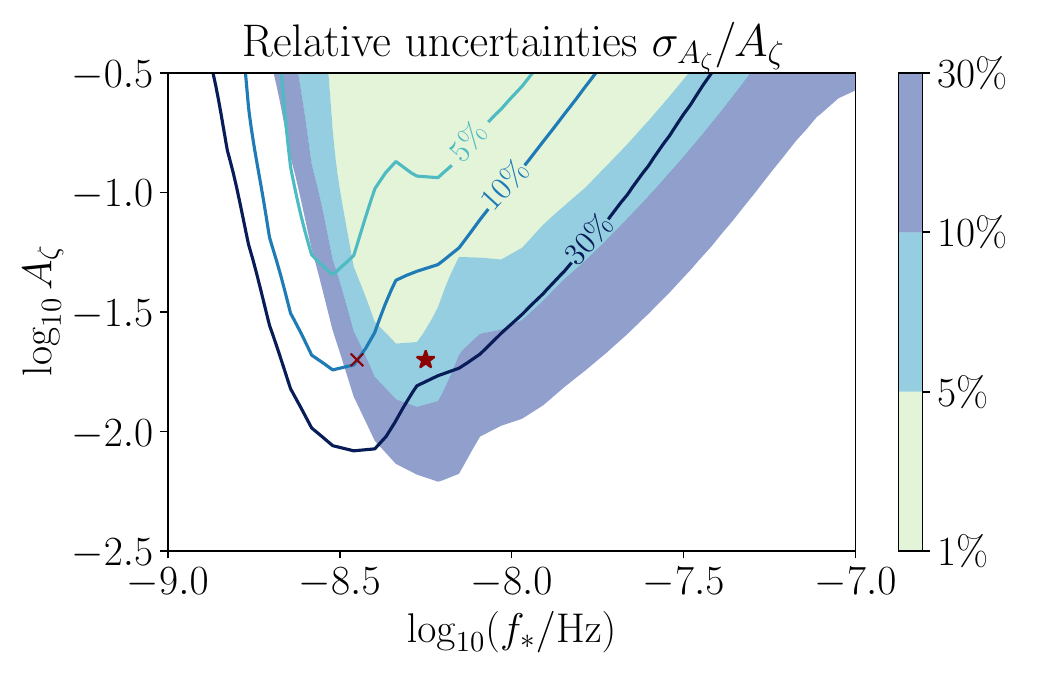}
\caption{
Same as Fig.~\ref{fig:Errors2D_Amplitude} but assuming 
$\aPL$ to be at the current best fit $+2\sigma$ (top row)
and $-2\sigma$ (bottom row).
}
\label{fig:Errors2D_Amplitude_2}
\end{figure*}

Focusing on the right panel, we see that the best reconstruction is obtained for a peak frequency around $f_* \sim \text{few} / T_{\rm obs}$, depending on the observation time we consider. For the current NANOGrav duration of $16.03$ yr, the peak sensitivity falls at $f_*\simeq 10^{-8.5}$Hz. The double peak feature reflects the shape of the SIGW spectrum for the assumed narrow width of ${\cal P}_\zeta$. The uncertainty quickly grows going away from the sweet spot values either because the signal falls outside the observable band (at low pivot frequencies) or it can only be observed above the pulsar noise in its tail (at high pivot frequencies).
As we already saw in the examples of the previous sections, a subdominant SIGW can be measured with a precision better than ${\cal O}(30)\%$ only in a limited region of parameter space. 

We validate the results of the FIM estimates by comparing them with MCMC Bayesian inference of simulated frequency domain datasets. When running MCMC analysis, we impose the following priors on the model parameters
\begin{equation}
    \theta \in 
    [\theta_0 - 5\, \sigma_\theta,
    \theta_0 + 5\, \sigma_\theta], 
\end{equation}
where the subscript 0 denotes the injection values, while $\sigma_\theta$ is the uncertainty estimated using the FIM analysis. The SIGW priors are further limited by the following ranges
\begin{align}
    \log_{10}\Az &\in [-3.5, \, 0], 
    \nonumber \\
    \log_{10}\Delta &  \in [-1.5,\, 0.9],
    \nonumber \\
    \log_{10}(f_*/{\rm Hz}) & \in [-10, \,-6].
\end{align}
Moreover, in the MCMC run we also impose a prior that avoids PBH overproduction (see next section), i.e., we impose $f_{\rm PBH}<1$, which partially cuts a portion of the parameter space with large $\Az$.

In Fig.~\ref{fig:mcmc_modelp}, we show the posterior distribution on the PL+SIGW analyses for some benchmark points identified in the plane of Fig.~\ref{fig:Errors2D_Amplitude} by crosses. Notice that since we are interested in comparing the measurement uncertainties from the MCMC procedure and with the FIM, in this case, we do not generate data by sampling from a Gaussian distribution with zero mean and standard deviation given by the corresponding spectral density. Rather, we fixed the data to match with the ensemble average. Effectively, this corresponds to anchoring the mean of the realization (and, correspondingly, the mean values of all the parameters) to its expectation value, which allows us to focus on the uncertainties.
%
%
In general, we find very good agreement between the results of the MCMC analysis and the uncertainties estimated through the FIM. 
In the left panel, a slight non-Gaussianity of the green MCMC posterior is found, with small tails of the distribution appearing at low values of $\aPL$, $\Az$, and $\Delta$. This is due to the relatively small SNR of the injected signal. 
Contrarily, in the right panel, we notice that the green contours appear to be slightly more constraining than the red ones. The origin of this small difference comes from the PBH overproduction prior, which is imposed in the MCMC, but not in the FIM. 
For this reason, we also show, in blue, the impact of including the same prior in the FIM analysis.\footnote{In practice, this is achieved by sampling data from the FIM and rejecting points that are not compatible with $f_{\rm PBH}<1 $.} After including the same prior, while this does not introduce major corrections to the posterior, we see the constraints to match nearly perfectly.

We repeat the exercise in Fig.~\ref{fig:Errors2D_Amplitude_2}, by varying the amplitude of the PL foreground within the $2\text{-}\sigma$ C.L. As expected, the uncertainties on $\Az$ are anti-correlated with the choice of injected $\aPL$. Overall, we conclude that varying the amplitude of the PL foreground within the $95\%$ C.L. results in a shift of around half an order of magnitude in the reach of future PTA datasets on $\Az$. This would be significant in terms of constraining power on PBH formation scenarios, as we motivate in the following section.

\begin{figure*}\centering
\includegraphics[width=0.49\textwidth, trim = 0cm 0cm 0cm 0cm]{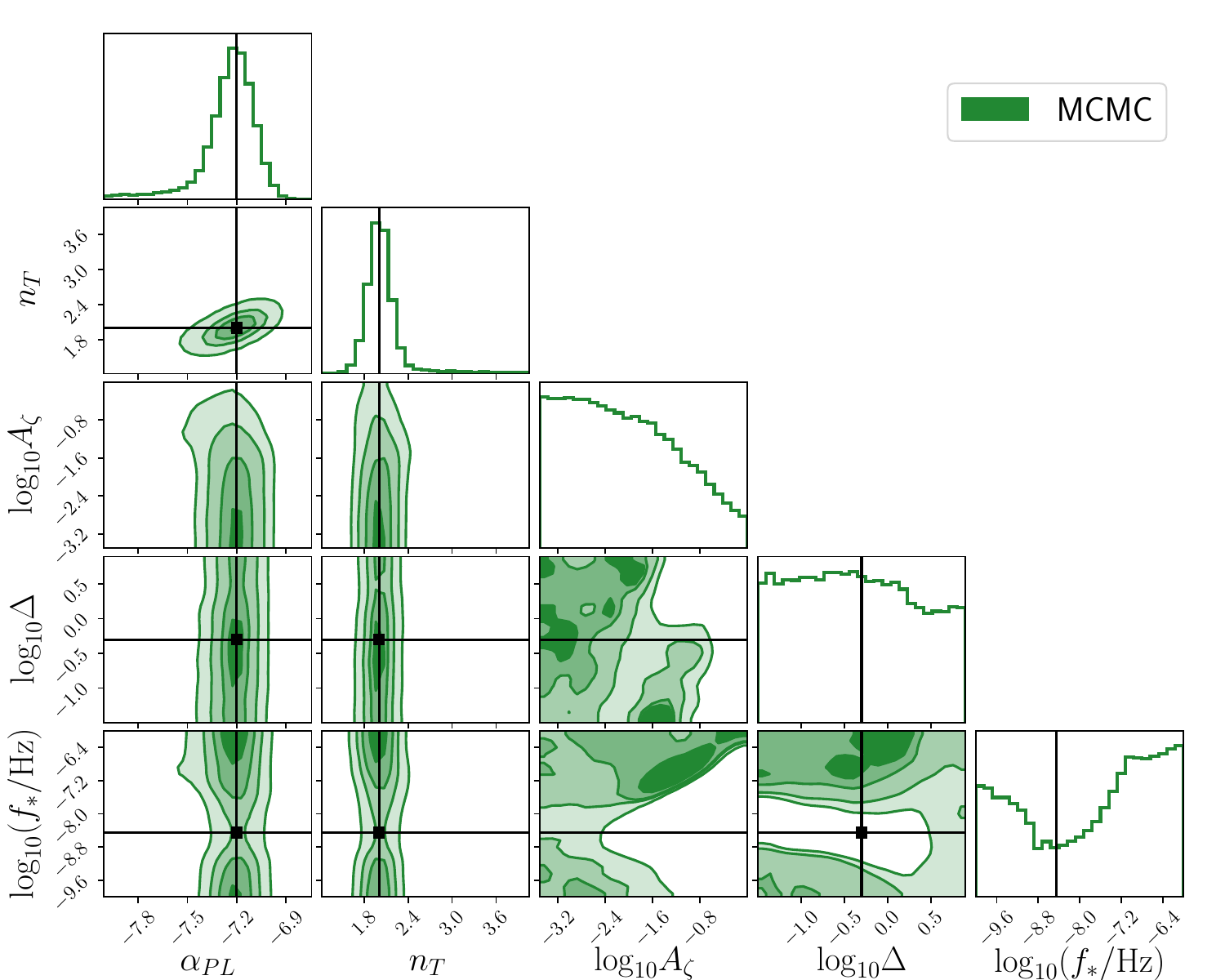}
\includegraphics[width=0.49\textwidth]{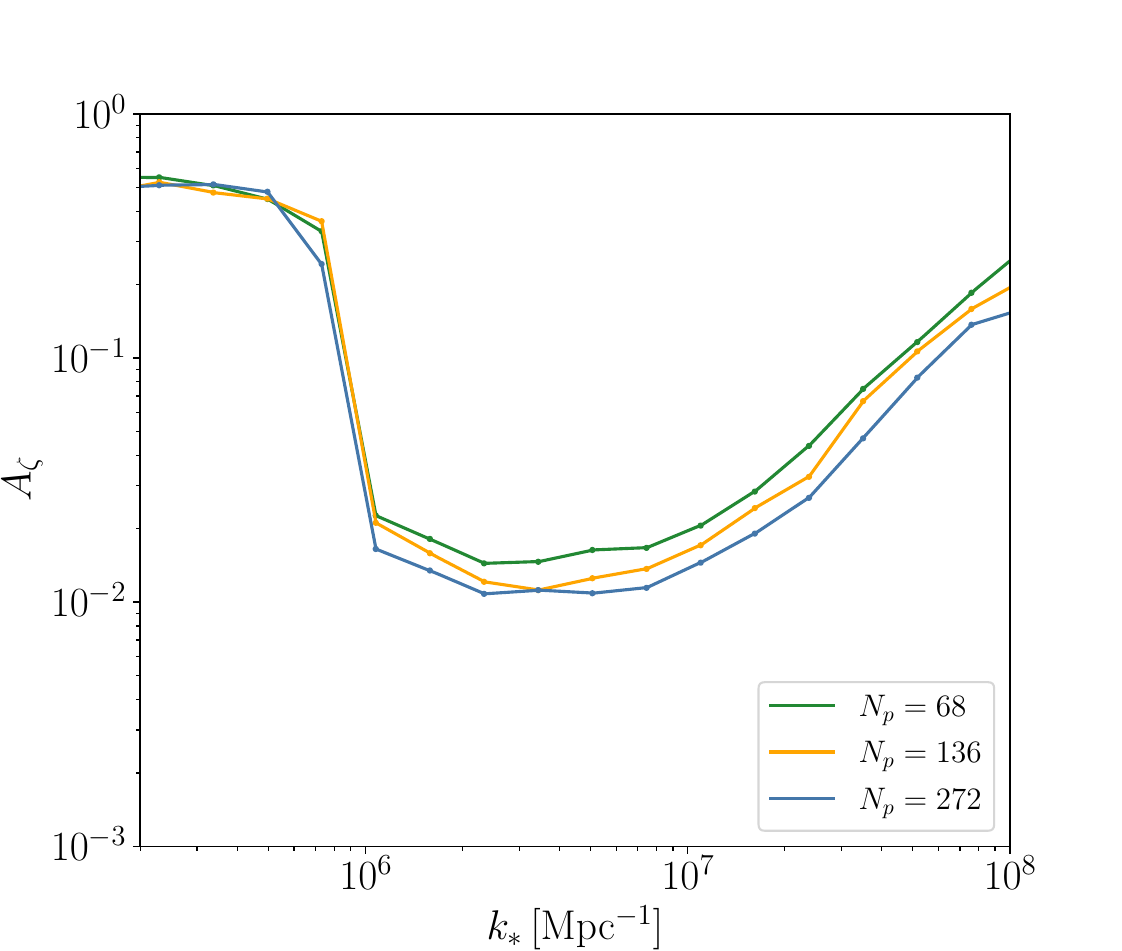}
\caption{
{\it Left panel:}
Posterior distribution obtained performing a Bayesian inference of the simulated power law signal as in Eq.~\eqref{eq:refPL} with a negligible SIGW contribution.
The dataset simulates a EPTA-like observation campaign with $T_{\rm obs} = 16.03\,$ yr, and $\Np = 272$. 
{\it Right panel:}
Zoom in the $(k_*,\Az)$ panel, reporting the upper bound on $A_{\zeta}$ at the $95$\% C.L., marginalizing over the spectral width and assuming $\Np = 68$ (\textit{green}), $\Np = 136$ (\textit{orange}), and $\Np = 272$ (\textit{blue}). 
}
\label{fig:Upper_bound_lowSIGW_2}
\end{figure*}

\section{Implications for PBHs}\label{sec:implications}

In this section, we explore the implications of a detection or upper bound on curvature perturbations for PBH models. 
Large primordial density perturbations can lead to the formation of a significant abundance of PBHs (see~\cite{LISACosmologyWorkingGroup:2023njw} for a recent review). 
Let us connect the scale of the perturbations, or equivalently the comoving wavenumber $k$, with the Hubble sphere mass 
$M_H$ as~\cite{Franciolini:2022tfm},
\begin{equation}
    M_H \simeq  
 10 M_\odot
 \left(\frac{g_*}{106.75}\right)^{1/2}
 \left(\frac{106.75}{g_{*s}}\right)^{2/3}
 \left(
 \frac{10^{6} {\rm Mpc}^{-1}}{k/\kappa}\right)^{2}.
 \label{M-k}
\end{equation}
The resulting PBH mass after the collapse is related to $M_H$ 
 by an order-unity factor described by the theory of critical collapse and validated through dedicated numerical relativistic simulations~\cite{Musco:2008hv} (see Ref.~\cite{Escriva:2021aeh} for a recent review).
Notice that we include the prefactor $\kappa \equiv k r_m $
that relates $k$ to the characteristic perturbation size at Hubble crossing $r_m = 1/a H$~\cite{Germani:2018jgr,Musco:2018rwt,Escriva:2019phb,Young:2019osy}, which also corresponds to the peak of the compaction function.
For definiteness, we will fix it to the value $\kappa = 2$ that corresponds to the case of a LN curvature spectrum with width 
$\Delta \simeq {\cal O}(0.5)$~\cite{Musco:2020jjb,Delos:2023fpm}.
Directly inspecting the right panel of Fig.~\ref{fig:Errors2D_Amplitude}, we see that PTA observations with $T_{\rm obs} = {\cal O} (10)\text{yr}$ have a peak sensitivity for masses around $M \simeq {\cal O}({\rm few})\, M_\odot $, with an extended tail in the subsolar mass range.

The PBH abundance can be computed by evaluating the probability that the compaction function, a measure of the mass excess over the areal radius, is larger than the threshold for collapse, which is defined through dedicated numerical simulations. We follow the general prescription presented in ref.\,\cite{Ferrante:2022mui} (see also refs.\,\cite{Young:2022phe,Gow:2022jfb}), where more details can be found.
The approach is based on threshold statistics on the compaction function and can include NG corrections to the statistics, although we only assume Gaussian perturbations in this work.
We review these computations in App.~\ref{app:PBHabundance} for completeness.
This is also built in an independent module of the \href{https://github.com/Mauropieroni/fastPTA/}{\texttt{fastPTA}} code and can be used to evaluate overproduction bounds when running MCMC Bayesian analysis both on simulated and real GW data.

In this section, we derive the upper bound on the spectral amplitude and compare it to the values required to produce a significant population of PBHs. 
Assuming no SIGW detection, an agnostic upper bound is obtained by injecting a PL signal with best-fit parameters and performing a multi-model fit which could include 
a subdominant SIGW signal. 
The priors we impose on the SIGW parameters are 
\begin{align}
\log_{10}A_{\zeta} &\in [-3.5, 0], 
\nonumber \\
\log_{10}\Delta    &\in [-1.5, 0.9], 
\nonumber \\
\log_{10}(f_*/\mathrm{Hz})) &\in [-10, -6].
\end{align} 
We do not include additional priors from PBH overproduction in this first stage.

In Fig.~\ref{fig:Upper_bound_lowSIGW_2}, we derive the upper bound on the power spectral parameters $\Az$ and $k_*$, marginalized over the power spectral width $\Delta$, assuming no SIGW is detected using the MCMC analyses. 
In the left panel, we show the full MCMC posterior distribution. We see that the PL parameters are well reconstructed, with bounds on $\aPL$ and $n_T$ which are competitive with analyses assuming a single PL. A slight degeneracy between the PL and SIGWs is observed, and a subdominant tail of the posterior explains the injected signal as coming from SIGWs, with $\Az \simeq 0.1$ and $\log_{10}(f_*/{\rm Hz})\simeq  -7.2$. This is also observed as a darker contour in the $(\Az, f_*)$ plane. For pivot scales away from the sweet-spot frequency, the shape of the SIGW becomes unconstrained, as can be seen by the flat posterior distribution of $\Delta$.

The plots depict a clear upper bound on $\Az$ as a function of the pivot scale, marginalized over the spectral width.
We focus on the $(\Az, f_*)$ plane again in the right panel, where we show marginalized upper bounds at 95\% C.L. for three different datasets, assuming $\Np = 68$ (\textit{green}), $\Np = 136$ (\textit{orange}), and $\Np = 272$ (\textit{blue}) of EPTA-like catalogs with $T_{\rm obs} = 16.03$ yr. 
The green line, corresponding to $\Np = 68$, is compatible with the current most stringent upper bound derived from existing NANOGrav data. This line serves as a benchmark for comparison, illustrating the present constraints on parameter uncertainties. The other lines correspond to projections based on future IPTA analyses or subsequent datasets, such as those anticipated from enhanced observations with larger PTA datasets. 
Slight oscillatory features observed in the plot are only due to the finite sampling of the posterior. 
These scenarios demonstrate a progressive tightening of the uncertainties, leading to stronger upper bounds.
Notice there exists a difference with the result reported in Fig.~2 (left panel) of~\cite{Iovino:2024tyg}, where a much tighter upper bound was derived from the currently available PTA dataset (to be compared with our green line $\Np = 68$).
The difference comes from our marginalization over $\Delta$, which was instead fixed to the representative value of $\Delta =0.5$ in~\cite{Iovino:2024tyg}.

\begin{figure}[t]\centering
\includegraphics[width=0.49\textwidth,trim = 0cm 0cm 0cm 1.2cm]{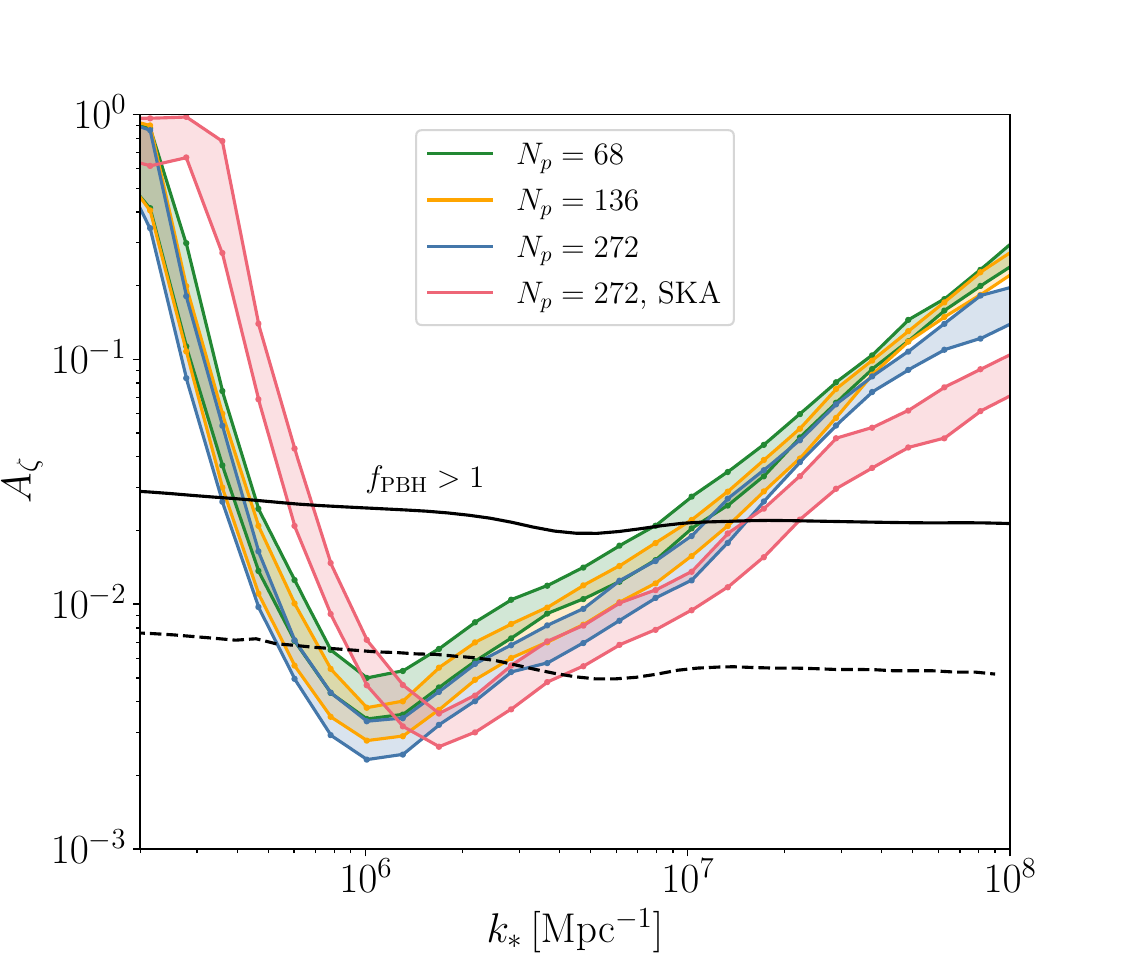}
\caption{
Same as Fig.~\ref{fig:Upper_bound_lowSIGW_2} (right panel), but fixing $\Delta = 0.5$. Additionally, the bound obtained from SKA-like data with $T_{\rm obs} = 10.33$ yr and $\Np = 272$ is reported in red. $T_{\rm  obs} = 16.03$ yr is assumed for EPTA-like catalogs with different numbers of pulsars. 
The band indicates the upper bound at $95$\% and $99.7$\% C.L.
The black solid and dashed lines indicates $f_{\rm PBH} = 1$
using threshold statistics or peak theory, respectively. In both cases, we assume Gaussian perturbations. 
}
\label{fig:Upper_bound_lowSIGW_3}
\end{figure}

To facilitate the comparison of the forecasted bounds with PBH abundance, as well as previous literature, we further restrict the width of the PBH spectrum to be $\Delta = 0.5$. 
In Fig.~\ref{fig:Upper_bound_lowSIGW_3}, we show the corresponding upper bound on the spectral amplitude as a function of the pivot scale.
The color code is the same adpoted in Fig.~\ref{fig:Upper_bound_lowSIGW_2} (right panel), with the 
colored band corresponding to $95$\% and $99.7$\% C.L.
We find that by fixing the width to the choice made in \cite{Iovino:2024tyg}, our estimate of the current PTA bound agrees with their derivation from the real NANOGrav data when assuming nearly Gaussian perturbations.\footnote{A minor difference is observed around $k_* \simeq 10^7 {\rm Mpc}^{-1}$, where the actual NANOGrav 15yr  bound is slightly more stringent.  As we work with simulated datasets, we are not able to reproduce real data statistical fluctuations or unmodeled systematics. Furthermore, it should be noted that those frequency bins only provide an upper bound on the SGWB amplitude, and the 95\% C.L. upper bound is thus still sensitive to the choice of the lower edge of the prior range for the SGWB amplitude.}  
We also estimate an SKA-like scenario with $N_p = 272$ and $T_{\rm obs} = 10.33$yr. The bound shifts to an higher pivot frequency due to the shorter observation time we assumed. It is interesting to stress, though, that the bound does not improve significantly moving from EPTA-like to SKA-like noise. This is a consequence of the fact that we enter in the signal-dominated regime with the PL dominating the SGWB in the observed frequency range.

We also superimpose a black solid line to show the amplitude corresponding to the representative value $f_{\rm PBH} = 1$ computed assuming threshold statistics. 
In the same plot, the black dashed line indicates the value of the amplitude computed assuming peak theory \cite{Bardeen:1985tr}. 
The slight reduction observed in both lines around $k\simeq 6 \times 10^6\, {\rm Mpc}^{-1}$ is due to the effect of the QCD era in reducing the threshold for collapse \cite{Jedamzik:1996mr,Byrnes:2018clq,Franciolini:2022tfm,Escriva:2022bwe,Musco:2023dak}. 
The difference between these two lines indicates the systematic theoretical uncertainties associated with this computation \cite{Green:2004wb,Young:2014ana,DeLuca:2019qsy,Frosina:2023nxu}. While threshold statistics predicts a larger amplitude for fixed abundance and reduces the tension with the SIGW explanation of the current PTA signal, it would also lead to tighter bounds on the PBH production in the range $k\in [4 \times 10^{5}, 10^7]\,{\rm Mpc}^{-1}$, which corresponds to masses in the range $M_{\rm PBH} \sim 0.6 M_H \in [0.28, 200] M_\odot$, right in the range of masses probed by the LVK collaboration \cite{Franciolini:2021xbq} and by future next-generation ground-based experiments \cite{DeLuca:2021hde,Pujolas:2021yaw,Franciolini:2023opt,Branchesi:2023mws}. 
Conversely, assuming peak theory, one obtains a smaller required amplitude, which exacerbates the tension with a possible SIGW interpretation of current PTA data, but at the same time corresponds to very weak bounds on PBH production, only currently scratching the $3\sigma$ significance around $M_{\rm PBH} \sim 0.6 M_H \in [10, 56] M_\odot$. This theoretical uncertainty should be addressed elsewhere in dedicated studies. 

It is striking that, although the dataset is assumed to grow by a factor of $4$ in terms of the number of pulsars,  the upper bound from SIGW only improves by less than a factor of $2$, due to the dominant role of the PL signal in the nHz band. 
We, therefore, foresee much of the progress in constraining PBH models coming from theoretical advancements in reducing {\it uncertainties} in the PBH computations, which should resolve the tension between the different methods.
Additionally, we do not address the additional {\it model dependence} introduced by possible non-Gaussianity of curvature perturbations, and refer to \cite{Iovino:2024tyg} for a thorough discussion.

\begin{figure}[t]\centering
\includegraphics[width=0.5\textwidth,
trim = 0cm 0cm 0cm 1.2cm]{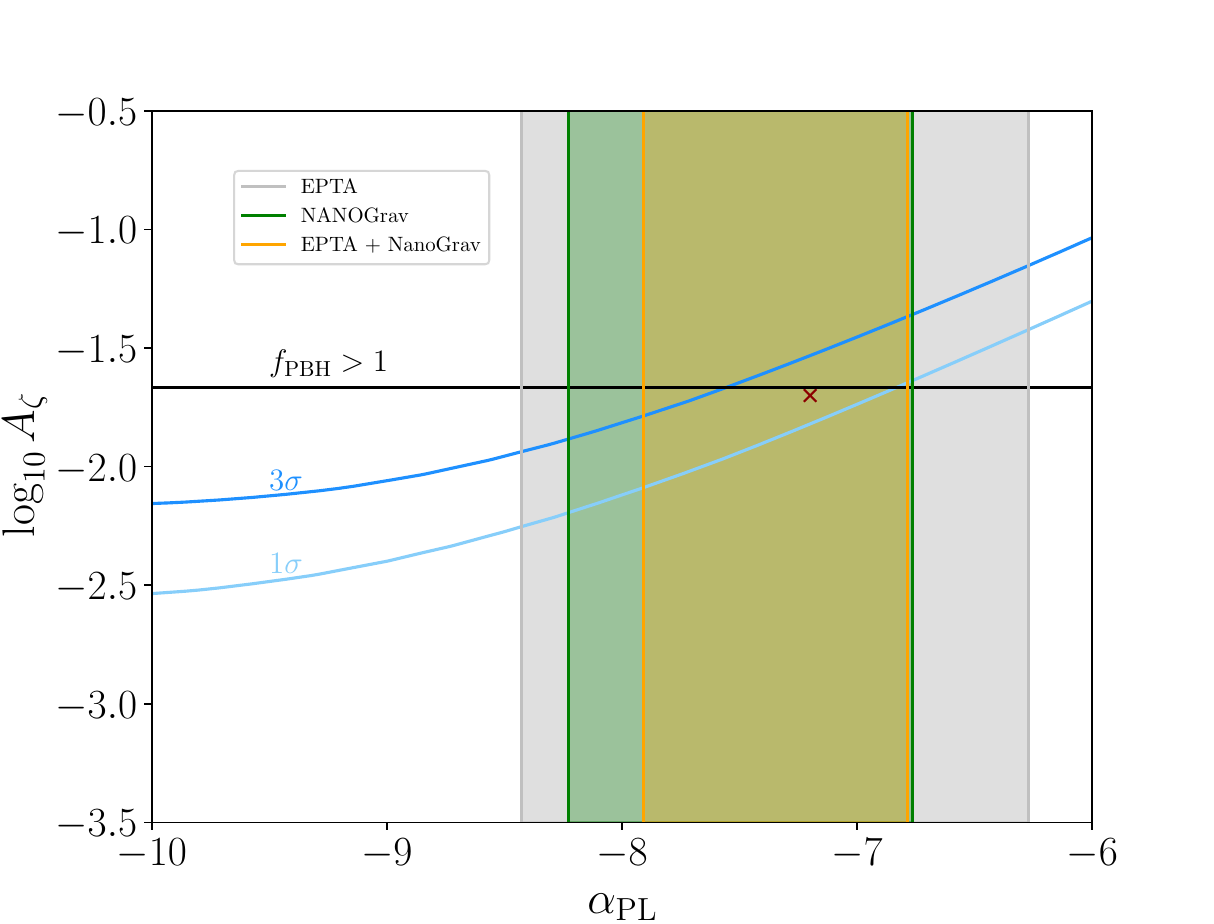}
\caption{
 $1 \sigma$ and $3 \sigma$ detectability level for $A_{\zeta}$ for a SKA-like configuration with fixed $T_{\rm obs} = 10.33$ yr and $N_{p} = 200$. 
The black horizontal line shows the PBH overproduction ($f_{\mathrm{PBH}} = 1$) amplitude assuming threshold statistics. 
The colored vertical bands are the 95\% C.L. for $\alpha_{\rm PL}$ as computed from EPTA data (gray), NANOGrav data (green), and EPTA + NANOGrav data (orange), see~\cite{Figueroa:2023zhu}. 
The other parameters are fixed as in Fig.~\ref{fig:SIGW_plot}.
The injection values for the MCMC analysis shown in Fig.~\ref{fig:mcmc_modelp} are marked with a red cross. 
}
\label{fig:Amplitude_range}
\end{figure}

It should be stressed that the conclusions above are subject to the assumption that the dominant PL signal is described by the current best-fit PL parameters \eqref{eq:refPL}. The actual future observational bound may also change depending on the assumptions we make on the true value of $\aPL$. 
We explore this dependence in Fig.~\ref{fig:Amplitude_range}, where we compare the FIM uncertainties on the power spectral amplitude to the upper bounds of the PBH abundance
while also varying the assumed amplitude of the foreground $\aPL$.
In blue, we indicate the upper bound at $1\sigma$  C.L., i.e. $\Delta A_{\zeta}/A_{\zeta} < 1$, and $3\sigma$  C.L., i.e. $\Delta A_{\zeta}/A_{\zeta} < 1/3$.
As expected, the upper bound on $\Az$ scales with the assumed value of $\aPL$. At the high end, we find ${\rm Max}(\Az) \sim \aPL^{1/3}$.
On the other hand, the upper bound flattens for sufficiently low values of $\aPL$, where the GW foreground becomes irrelevant and the bound is driven by the pulsar noises. 
The SIGW parameters not shown are fixed to ${\rm log}_{10} \Delta = -0.3$ and ${\rm log}_{10} (f_*/{\rm Hz}) = -7.8 $, as in Fig.~\ref{fig:SIGW_plot}.
Most notably, varying $\aPL$ within the current most stringent 95\% C.L. reported by the NANOGrav collaboration (green band), results in a shift of the upper bound on $\Az$ by around half an order of magnitude. This range is compatible with the current systematic uncertainties on the computation of the PBH abundance. 
Therefore, further reduction of the uncertainties expected with future datasets will have to be accompanied by a drastic reduction of the theoretical uncertainties associated with the PBH abundance computation.

\section{Conclusions}\label{sec:conclusions}

In this work, we have explored the future constraints that PTA observations could impose on SIGWs and the associated primordial curvature power spectrum. We mostly employed an FIM approach, which was validated through specific full Bayesian MCMC runs, and we evaluated the precision with which upcoming PTA datasets -- characterized by increased numbers of pulsars, extended observation times, and lower white noise-- will be able to probe SIGWs.

Our analysis indicates that, while current PTA data is still plagued by significant uncertainties on the SIGW properties, future PTA observations with larger pulsar networks will drastically improve sensitivity. This will lead to reduced degeneracies between key parameters, such as the amplitude, width, and peak frequency of the SIGW spectrum, ultimately providing stronger constraints on the primordial power spectrum.
In case SIGW is the sole responsible for the PTA excess, future PTA datasets will slowly increase precision in the determination of the spectral amplitude and tilt, with uncertainties remaining relatively large due to the necessarily low signal amplitude forced by the PBH overproduction bound. 
Efforts should be directed towards testing this hypothesis using information beyond the frequency spectrum.
Moreover, in the scenario in which SIGW only accounts for a subdominant contribution to current data, 
we have investigated how future datasets could improve our constraints on PBH formation. While current uncertainties limit precise determinations of the PBH abundance, the enhanced sensitivity of the next PTA datasets will either impose stringent upper bounds on PBH populations or detect signatures of primordial density fluctuations responsible for their formation.
We stressed how as measurement uncertainties narrow down, 
the theoretical uncertainties on the determination of the PBH abundance could remain a limiting factor, potentially preventing us from drawing strong conclusions.

The potential existence of an astrophysical foreground, i.e. sourced by SMBH mergers, remains a key factor in SIGW detectability. 
A reduction of the uncertainties on the possible foreground amplitude, i.e. leading to a more precise determination of the GW foreground to be expected, 
would allow a more precise determination of the future upper bound on SIGWs.

We conclude by mentioning a few caveats underlying our analyses, which could be independently addressed in future work. 
The present analysis relies solely on the information provided by the spectral shape of different SGWB contributions without identifying any smoking-gun feature distinguishing astrophysical from cosmological sources. While this Bayesian model comparison approach is standard in GW data interpretation, the conclusions drawn remain inherently dependent on the reliability of the assumed models. In this respect, the template employed for the curvature power spectrum is relatively flexible, yet the second-order nature of the induced signal imposes constraints on its shape. A more robust modeling of the astrophysical sources would refine these constraints and enhance their significance. Moreover, an explicit model comparison incorporating a first-principles computation of ${\cal P}_\zeta$ within an early-universe framework would yield even more stringent bounds, as discussed in~\cite{LISACosmologyWorkingGroup:2025vdz}.

In our study, we assumed that the dominant astrophysical foreground follows a power-law spectrum. Future observations with greater sensitivity might reveal deviations from this assumption, manifesting as spectral oscillations absent from the SIGW signal. Such deviations could improve the differentiation between astrophysical and cosmological components. Additionally, we conservatively considered the astrophysical SGWB to be irreducible through the subtraction of individually resolved sources. However, in practice, improved source subtraction techniques might reduce its contribution, thereby further clarifying the potential SIGW signature.

Another distinguishing aspect is that the astrophysical foreground exhibits significantly stronger anisotropies, particularly at high frequencies, which are instead not present in the cosmological SIGW at large scales~\cite{Bartolo:2019zvb}. Once detected, incorporating these anisotropies into the analysis could improve the separation of the subdominant cosmological SGWB component. The study of these anisotropic effects is left for future research.

To extend the scope of our analysis, we envision the use of our forecast code in conjunction with \href{https://github.com/jonaselgammal/SIGWAY}{\texttt{SIGWAY}}. This combination would enable the exploration of more model-independent spectral parameterizations and facilitate the direct modeling of the inflationary dynamics responsible for large curvature perturbations.
Additionally, one should explore more in detail deviations from the perfect radiation assumption adopted here, as would QCD thermal history unavoidably affect the PTA scale~\cite{Schwarz:1997gv,Watanabe:2006qe,Franciolini:2023wjm}, as well as possible new physics beyond the standard model.  We leave this for future work.

\begin{acknowledgments}
 We thank Valerie Domcke, Jonas El Gammal, Antonio Iovino, Gabriele Perna, Nataliya Poryako, Antonio Riotto, and the EPTA GW interpretation group for interesting discussions. 
CC thanks the TH department at CERN for the kind hospitality during the beginning of this project. 
MP acknowledges the hospitality of
Imperial College London, which provided office space during some parts of this project. CC acknowledges support from the Istituto Nazionale di Fisica Nucleare (INFN) through the Commissione Scientifica Nazionale 4 (CSN4) Iniziativa Specifica “Quantum Fields in Gravity, Cosmology and Black Holes” (FLAG).
\end{acknowledgments}

\appendix 

\section{Computing the PBH abundance}\label{app:PBHabundance}

One can compute the PBH abundance by integrating the mass fraction $\beta(M_H)$ produced when perturbations re-enter the Hubble sphere at all relevant scales, including the appropriate redshift scaling. 
This can be computed as ~\cite{Sasaki:1986hm}
\begin{align}
  f_{\rm PBH} &= 
  \frac{1}{\Omega_{\text{DM}}} 
  \int_{M_{H,{\rm min}}}
  ^{M_{H,{\rm max}}}
\frac{{\rm d} M_H}{M_H} 
\left( \frac{1}{M_H} \right)^{1\over2} 
\nonumber \\
& \times 
\left( \frac{g_*}{106.75} \right)^{3\over 4} 
\left( \frac{g_{*S}}{106.75} \right)^{-1} 
\left( \frac{\beta(M_H)}{7.9 \times 10^{-10}} \right), 
\, 
\label{beta}
\end{align}
where we introduced the fraction of PBHs 
compared to the DM budget $f_{\rm PBH} \equiv \Omega_{\rm PBH}/\Omega_{\rm DM}$.
Furthermore, the effective degrees of freedom should be evaluated at the corresponding epoch, identifying the Hubble mass with the temperature using
\begin{equation}
    M_H(T_k) 
    = 4.8\times 10^{-2} M_\odot 
    \left(\frac{g_*}{106.75} \right)^{-{1 \over 2}} \left( T_k\over{\rm GeV} \right)^{-2}.
\end{equation}

Given Eq.~\eqref{beta}, computing the abundance boils down to evaluating the probability a given Hubble patch collapses to form a PBH. 
We will follow the general prescription presented in ref.\,\cite{Ferrante:2022mui} (see also refs.\,\cite{Young:2022phe,Gow:2022jfb}), where more details can be found.
The approach is based on threshold statistics on the compaction function, a fundamental variable defined as twice the local mass $\delta M(r,t)$ excess over the areal radius $R(r,t)$
\begin{align}\label{eq:DefinitionCompaction}
\mathcal{C}(r,t) 
\equiv
\frac{2\delta M(r,t)}{R (r,t)} 
= 
\frac{2}{R(r,t)}
\int_{S^2_R} d^{3}\vec{x}\,\rho_b(t)\,\delta(\vec{x},t)\,.
\end{align}
Assuming spherical symmetry and adopting the gradient expansion approximation, on the super-Hubble scale the relation between the density contrast $\delta$ and the curvature perturbation $\zeta$ is non-linear\,\cite{Harada:2015yda,Musco:2018rwt} 
\begin{align}\label{eq:SphericalDelta}
\delta(r,t) = 
-\frac{2 \Phi}{3}
\left(\frac{1}{aH}\right)^2 
e^{-2\zeta(r)}\left[
\zeta^{\prime\prime}(r) + \frac{2}{r}\zeta^{\prime}(r) + \frac{1}{2}\zeta^{\prime}(r)^2
\right]\,.
\end{align}
The parameter $\Phi$ is introduced to keep track of how the equation of state of the Universe changes due to the thermal history, which is particularly relevant for the formation of stellar mass PBH across the QCD phase transition.

As the relation between the density contrast and curvature perturbations is non-linear, the above equation unavoidably introduces a certain amount of NGs in the PDF of $\delta$, and hence on that of $\mathcal{C}$. Integrating Eq.~\eqref{eq:SphericalDelta} over the radial coordinate, one finds 
\begin{align}\label{eq:CompactionFull}
\mathcal{C}(r) = 
-2\Phi\,r\,\zeta^{\prime}(r)\left[
1 + \frac{r}{2}\zeta^{\prime}(r)
\right] = 
\mathcal{C}_1(r) - \frac{1}{4\Phi}\mathcal{C}_1(r)^2\,,
\end{align} 
where $\mathcal{C}_1(r) \equiv -2\Phi\,r\,\zeta^{\prime}(r)$ is defined as the linear component of the compaction function. 
In full generality, the latter can be recast in terms of its Gaussian counterpart as follows
\begin{align}
    \mathcal{C}_1(r) = \mathcal{C}_{\rm G}(r)\,\frac{d F}{d\zeta_{\rm G}} \, ,~~~~~~~ \mathcal{C}_G(r) \coloneqq -2\Phi\,r\,\zeta^{\prime}_G(r)\,,
\end{align}
where $F$ encodes the relation between $\zeta$ and the Gaussian component $\zeta_{\rm G}$.

We stress that $\mathcal{C}(r)$ depends both on $\zeta_{\rm G}$ and $\mathcal{C}_{\rm G}$ which are, by definition, Gaussian distributed.
Therefore, their joint PDF can be computed as 
\begin{align}
\mathrm{P}_{\mathrm{G}}\left(\mathcal{C}_{\rm G}, \zeta_{\mathrm{G}}\right)
=
\frac{1}{2\pi\sigma_{c} \sigma_{r} \sqrt{1-\gamma_{c r}^{2}}} \exp \left(-\frac{\zeta_{\mathrm{G}}^{2}}{2 \sigma_{r}^{2}}\right) 
\nonumber 
\\
\times \exp \left[-\frac{1}{2\left(1-\gamma_{c r}^{2}\right)}\left(\frac{\mathcal{C}_{\mathrm{G}}}{\sigma_{c}}-\frac{\gamma_{c r} \zeta_{\mathrm{G}}}{\sigma_{r}}\right)^{2}\right],
\label{eq:PDFCG}
\end{align} 
where $\gamma_{cr}\equiv\sigma_{cr}^2/\sigma_c \sigma_r$ and the correlators are given by
 \begin{align}
 \sigma_c^2 &=
  \frac{4\Phi^2}{9}\int_0^{\infty}\frac{dk}{k}
  (kr_m)^4 W^2(k,r_m) P^T_{\zeta}(k)\,,\label{eq:Var1}
\nonumber \\
 \sigma_{cr}^2 & = 
 \frac{2\Phi}{3}\int_0^{\infty}\frac{dk}{k}(kr_m)^2
 W(k,r_m)
 W_s(k,r_m) P^T_{\zeta}(k)\,,
  \nonumber  \\
  \sigma_r^2 & =   \int_0^{\infty}\frac{dk}{k}
  W_s^2(k,r_m) P^T_{\zeta}(k)\,,
 \end{align}
 with 
 $P^T_{\zeta}(k) \equiv T^2(k,r_m) P_{\zeta}(k)$, 
 $W_s(k,r) = \sin(kr)/kr$ 
 while $W(k,R)$ and $T(k,\tau)$ are given by
 \begin{align}
 T(k,\tau) &= 3
 \left [
\frac{\sin(k\tau/\sqrt{3}) - (k\tau/\sqrt{3})\cos(k\tau/\sqrt{3})}{(k\tau/\sqrt{3})^3}
\right]\,,
\nonumber  \\
W(k,R) &= 3
\left[
\frac{\sin(kR) - kR\cos(kR)}{(kR)^3}
\right]\,.
\label{eq:T}
 \end{align}
The transfer function $T(k,\tau)$ we adopt in this section is derived assuming a perfect radiation fluid. 
Furthermore, we have selected the Hubble crossing epoch by fixing the conformal time $\tau = 1/ (a H) = r_m$, using the Hubble crossing condition $a H r_m = 1$.

Finally, $\beta(M_H)$ entering in Eq.~\eqref{beta}, which is identified with the mass fraction derived from the collapse of a single mode is 
\begin{align}
\beta(M_H) & = \int_{\mathcal{D}}\mathcal{K}(\mathcal{C} - \mathcal{C}_{\rm th})^{\gamma}
\textrm{P}_{\rm G}(\mathcal{C}_{\rm G},\zeta_{\rm G})d\mathcal{C}_{\rm G} d\zeta_{\rm G}\,,\label{eq:CompactionIntegral}
\end{align}
with the collapse condition being expressed as 
\begin{align}
 \mathcal{D} & = 
\left\{
\mathcal{C}(\mathcal{C}_{\rm G},\zeta_{\rm G}) > \mathcal{C}_{\rm th}  
\land\mathcal{C}_1(\mathcal{C}_{\rm G},\zeta_{\rm G}) < 2\Phi
\right\}\,.\label{eq:RegionD}
\end{align} 
The first condition requires overthreshold perturbations, while the second allows us only to consider type I PBH formation. 

Throughout this work, we have assumed Gaussian curvature perturbations. In this case, $F\equiv 1$ and one can marginalize over $\zeta_G$ in Eq.~\eqref{eq:PDFCG}, being the compaction function independent of $\zeta_G$.
Therefore, \eqref{eq:PDFCG} simply becomes a Gaussian distribution with variance $\sigma_c^2$. Notice that non-linearities induced by general relativity are still accounted for through the full non-linear compaction function entering the threshold condition in Eq.~\eqref{eq:CompactionIntegral}.

\bibliography{main}
\end{document}